\documentclass{article}
\usepackage[colorlinks]{hyperref}
\usepackage{amsmath,amssymb}
\usepackage{physics}
\usepackage{subcaption}
\usepackage{mathtools}
\usepackage{parskip}
\usepackage{graphicx}
\usepackage{comment}
\usepackage{bm}
\usepackage{amsthm}
\usepackage{upgreek}
\usepackage{tikz}
\usepackage[toc, page, title]{appendix}
\usepackage{algorithm}
\usepackage{algpseudocode}
\usepackage{authblk}

\usetikzlibrary{patterns,arrows,decorations.pathreplacing,math,decorations.markings,angles}

\providecommand{\keywords}[1]
{
	\small	
	\textbf{Keywords. } #1
}

\providecommand{\classification}[1]
{
	\small	
	\textbf{MSC subject classifications. } #1
}

\NewDocumentCommand{\expect}{ e{^} s o >{\SplitArgument{1}{|}}m }{%
	\operatorname{\mathbb{E}}
	\IfValueT{#1}{{\!}^{#1}}
	\IfBooleanTF{#2}{
		\expectarg*{\expectvar#4}%
	}{
		\IfNoValueTF{#3}{
			\expectarg{\expectvar#4}%
		}{
			\expectarg[#3]{\expectvar#4}%
		}%
	}%
}
\NewDocumentCommand{\expectvar}{mm}{%
	#1\IfValueT{#2}{\nonscript\;\delimsize\vert\nonscript\;#2}%
}
\DeclarePairedDelimiterX{\expectarg}[1]{[}{]}{#1}

\NewDocumentCommand{\variance}{ e{^} s o >{\SplitArgument{1}{|}}m }{%
	\operatorname{\mathbb{V}}
	\IfValueT{#1}{{\!}^{#1}}
	\IfBooleanTF{#2}{
		\expectarg*{\expectvar#4}%
	}{
		\IfNoValueTF{#3}{
			\expectarg{\expectvar#4}%
		}{
			\expectarg[#3]{\expectvar#4}%
		}%
	}%
}

\tikzset{->-/.style={decoration={
			markings,
			mark=at position #1 with {\arrow{>}}},postaction={decorate}}}
\newcommand{\AxisRotator}[1][rotate=0]{%
	\tikz [x=0.25cm,y=0.60cm,line width=.2ex,-stealth,#1] \draw (0,0) arc (-150:150:0.5 and 0.5);%
}

\algnewcommand{\Input}[1]{%
	\State \textbf{Input:}
	\Statex \hspace*{\algorithmicindent}\parbox[t]{.8\linewidth}{\raggedright #1}
}
\algnewcommand{\Output}[1]{%
	\State \textbf{Output:}
	\Statex \hspace*{\algorithmicindent}\parbox[t]{.8\linewidth}{\raggedright #1}
}

\title{Kinetic-Diffusion-Rotation Algorithm for Dose Estimation in Electron Beam Therapy}
\author[1]{Klaas Willems\thanks{klaas.willems@kuleuven.be}}
\author[1]{Vince Maes\thanks{vince.maes@kuleuven.be}}
\author[1]{Zhirui Tang\thanks{zhirui.tang@kuleuven.be}}
\author[1]{Giovanni Samaey\thanks{giovanni.samaey@kuleuven.be}}
\affil[1]{Department of Computer Science, KU Leuven, Leuven, Belgium}

\begin{document}

\maketitle

\begin{abstract}
    Monte Carlo methods are state-of-the-art when it comes to dosimetric computations in radiotherapy. However, the execution time of these methods suffers in high-collisional regimes. We address this problem by introducing a kinetic-diffusion particle tracing scheme. This algorithm, first proposed in the context of neutral transport in fusion energy, relies on the explicit simulation of the exact kinetic motion in low-collisional regimes and dynamically switches to an approximate random walk in high-collisional regimes. The random walk corresponds to an advection-diffusion process that preserves the first two moments (mean and variance) of the kinetic motion. We derive an analytic formula for the mean kinetic motion and discuss the addition of a multiple scattering distribution to the algorithm. In contrast to neutral transport, the electron beam therapy setting does not readily admit to an analytical expression for the variance of the kinetic motion, and we therefore resort to the use of a lookup table. We test the algorithm for dosimetric computations in electron beam therapy on a 2D CT scan of a lung patient. Using a simple particle model, our Python implementation of the algorithm is nearly 33 times faster than an equivalent kinetic simulation at the cost of a small modeling error. 
\end{abstract}
\keywords{Monte Carlo, electron beam radiation therapy, kinetic-diffusion}\\
\classification{35Q20, 60G50, 65C05, 92-10}

\newpage
\section{Introduction}
\label{section:intro}
Radiation therapy \cite{jabbari_review_2011} is a method for treating cancer. Malicious tissue is radiated by a beam of charged particles such as photons or electrons. These charged particles interact with human tissue and damage the DNA of cells. Cells whose DNA is damaged beyond repair stop dividing or die, destroying or shrinking the tumor in the process. Medical experts design patient-specific treatment plans based on a whole host of parameters such as the size of a tumor, depth, and location of the tumor, proximity of the tumor to organs, and so on \cite{rhee_automated_2020}. Often the treatment planning process involves radiation simulations using CT scans of the patient \cite{reynaert_monte_2007}. These simulations are used to optimize parameters such as the radiation time, radiation beam width, and radiation beam orientation \cite{kisling_fully_2019}. Within this optimization problem, the goal is to maximize the dose (deposited energy) to the tumor, while insisting the dose to the surrounding organs does not exceed a predefined amount. In this text, we focus exclusively on electron beam radiation therapy. The equation that governs the behavior of the high energy electrons in human tissue is a linear Boltzmann equation \cite{cercignani_boltzmann_1988} \begin{align}
	\label{eq:linearBoltzmann}
	\pdv{\varphi(\mathbf{x}, \mathbf{v}, t)}{t} + \mathbf{v} \cdot \grad_{\mathbf{x}} \varphi(\mathbf{x}, \mathbf{v}, t) = \mathcal{L}(\varphi),
\end{align} with $\mathbf{x} \in \mathbb{R}^3$ the position, $\mathbf{v} \in \mathbb{R}^3$ the velocity, $t \in \mathbb{R}_{\geq 0}$ the time, $\varphi(\mathbf{x}, \mathbf{v}, t)$ the particle distribution and $\mathcal{L}(\varphi)$ a collision operator that is specified in section \ref{section:radiation}. 

Many applications such as rarefied gas modeling \cite{boyd_hybrid_2012}, radiation transport \cite{al-beteri_designing_1993} and neutral transport in fusion \cite{maes_hilbert_2023, larsen_asymptotic_1974} are concerned with the Boltzmann equation. The Boltzmann equation \eqref{eq:linearBoltzmann} can be solved in several different ways. In many circumstances, deterministic grid-based methods are not a viable option \cite{garrett_comparison_2013}. Since the Boltzmann equation is high-dimensional, large grids are required to reach the desired accuracy. In most practical applications, the Boltzmann equation is therefore solved using a Monte Carlo approach, in which the underlying particle dynamics are simulated \cite{jabbari_review_2011}. The benefit of using Monte Carlo simulations is that, unlike in grid-based methods, the cost does not scale exponentially with the dimension, the so-called curse of dimensionality \cite{lovbak_accelerated_2023}. In addition, Monte Carlo algorithms are easily parallelizable and deal well with irregular domains, complex particle dynamics, and multi-species systems. The downside is the introduction of a statistical error \cite{vassiliev_monte_2017}, which decreases as $\mathcal{O}(N^{-0.5})$ as the number of particles $N$ is increased. Even more, when the collision rate of the particles is large (stiff collision operator), explicit simulations of the particle dynamics are not always computationally feasible \cite{mortier_kinetic-diffusion_2022}. In the context of electron beam therapy, depending on the type of particle and the density of the background medium, the collision rate of the particle is nearly always large \cite{olbrant_models_2012}. 

In electron beam therapy, the problem due to high collisionality is dealt with in several different ways. In so-called condensed history algorithms, particle tracks are divided into segments. Rather than sample many individual collisions, the net effect for each segment is sampled from a so-called multiple-scattering distribution \cite{seltzer_electron-photon_1991}, that are derived from approximate scattering models. Modern codes such as Macro Monte Carlo (MMC) \cite{neuenschwander_macro_1992, jacqmin_su-d-218-06_2012, ding_first_2006, neuenschwander_mmc-high-performance_1995} take this idea one step further. In a pre-processing step, detailed distributions of the state of a particle (energy, position, velocity, secondary particles, ...) are precomputed using full Monte Carlo simulations. These distributions are then used to transport particles in large macroscopic steps. Superposition Monte Carlo (SMC) \cite{keall_superposition_1996, keall_super-monte_1996} introduced the concept of track repetitions. Here, several thousands of tracks are pre-computed and stored in a lookup table. During actual simulation, scattering angles, energy loss, and path lengths from the pre-computed tracks are modified based on the local properties of the background tissue. Voxel-based Monte Carlo (VMC) \cite{kawrakow_3d_1996} builds upon the same idea. Instead of continuously reusing the same particle tracks from the lookup table, multiple particles are simulated together. This allows for re-using material-independent computations and random numbers on the fly. 

For the linear Boltzmann equation used in e.g. fusion energy, an alternative approach, the kinetic-diffusion Monte Carlo algorithm, was developed to alleviate some of the problems associated with high collisionality \cite{mortier_kinetic-diffusion_2022}. The kinetic-diffusion Monte Carlo (KDMC) algorithm is based on the observation that in a high-collisional regime, an advection-diffusion equation becomes a valid approximation to the governing linear Boltzmann equation \cite{maes_hilbert_2023}. In KDMC, the dynamics is sped up by using the kinetic motion only up to the first collision, after which the associated particle dynamics of the advection-diffusion equation is used for the remainder of the step. In addition, KDMC is asymptotic preserving, meaning that the algorithm is consistent in the limit of infinite collisionality and vanishing time step. For a finite time step, KDMC introduces a small bias \cite{mortier_advanced_2020}. However, this bias can be removed using Multilevel Monte Carlo (MLMC) algorithms \cite{mortier_advanced_2020, loevbak_multilevel_2023}. 

In this paper, we adapt the kinetic-diffusion scheme to simulations in electron beam therapy. We refer to the adapted version of KDMC as kinetic-diffusion-rotation (KDR). In section \ref{section:radiation}, we present the kinetic equation and associated particle model relevant to electron beam therapy. To motivate the required adaptations, we additionally highlight the differences and similarities between the linear Boltzmann equation for neutral transport in fusion energy, and the transport equation used in electron beam therapy. In section \ref{section:KD}, we introduce the novel KDR scheme, and describe the differences and similarities with respect to the KDMC scheme for neutral transport. In section \ref{section:numerical}, we apply KDR to a practical electron beam therapy simulation to show the potential of kinetic-diffusive-type algorithms. Finally, in section \ref{section:conclusion}, we present our conclusions. This text is a summary of the first author's master thesis \cite{willems_particle_2023}.

\section{Kinetic equation for electron beam therapy}
\label{section:radiation}
In section \ref{section:radiationModel}, we introduce the kinetic equation and associated particle model relevant to electron beam radiation therapy. For simplicity, we stick to a simplified electron model. In section \ref{section:fusionComparison}, we highlight the differences between linear Boltzmann equation and the transport equation used in electron beam therapy. This will provide insight into how the KDMC algorithm, originally developed in the context of fusion energy, can be adapted for use in electron beam simulations. 

\subsection{Kinetic equation for electron beam therapy}
\label{section:radiationModel}
When electrons travel through matter, they interact with it in various ways. The interactions relevant to electron beam therapy are elastic scattering, soft inelastic scattering, hard inelastic scattering, and bremsstrahlung \cite{olbrant_models_2012}. The design and testing of software that models all types of scattering accurately is out of the scope of this text. Instead, we consider a simplified electron model, derived from a realistic one. To this end, bremsstrahlung interactions and hard inelastic collisions are neglected. Soft elastic collisions are simulated explicitly. Finally, angular deflections due to soft inelastic scattering are neglected, and energy loss is deposited using the continuous slowing-down approximation \cite{vassiliev_monte_2017}. This is the simplest model that still contains all mathematical challenges related to the extension of KDMC to scattering-type collisions in electron beam radiation therapy. Generalizations to more complex particle models for kinetic-diffusion-type algorithms are discussed in section \ref{section:KD}. The kinetic equation that describes this simplified particle is the Boltzmann continuous slowing down (CSD) equation \cite{kupper_models_2016} \begin{multline}
	\label{eq:BoltzmannCSD}
	\underbrace{-\pdv{}{E} \left[S(E, \textbf{x})\psi(E, \textbf{x}, \Omega)\right]}_{\substack{\text{energy loss due to soft} \\ \text{inelastic collisions}}} + \underbrace{ \vphantom{\pdv{}{E}} \Omega \grad_{\textbf{x}} \psi(E, \textbf{x}, \Omega)}_{\text{transport}} = \\ \underbrace{\int_{\mathrm{S}^2} \Sigma_s(E, \textbf{x}, \Omega \cdot \Omega') \psi(E, \textbf{x}, \Omega') d\Omega'}_{\text{source due to soft elastic scattering}} - \underbrace{ \vphantom{\int_{\mathrm{S}^2}} \Sigma_t (E, \textbf{x}) \psi(E, \textbf{x}, \Omega)}_{\substack{\text{sink due to soft} \\ \text{elastic scattering}}}.
\end{multline} The CSD Boltzmann equation \eqref{eq:BoltzmannCSD} expresses the evolution of the population of electrons $\psi(E, \textbf{x}, \Omega)$ as they interact with the tissue as a function of the particles' energy $E \in \mathbb{R}$, position $\mathbf{x} \in \mathbb{R}^3$ and orientation (normalized velocity) $\Omega \in \mathbb{S}^2$. The left-hand side of equation \eqref{eq:BoltzmannCSD} describes the movement of particles through space as a particle's energy decreases. The rate at which the energy of a particle decreases is given by the stopping power $S(E, \textbf{x})$. Throughout this work we use the stopping power for distant and close interactions from PENELOPE \cite[eq. (3.120)]{nuclear_energy_agency_penelope_2019}. The right-hand side of equation \eqref{eq:BoltzmannCSD} contains a source term and a sink term, that together can be interpreted as a collision operator: Particles are removed from the population at a rate $\Sigma_t (E, \textbf{x})$ (sink), and are then immediately returned to the population with a new orientation through the integral term (source). Particles are spawned from an initial condition $\psi(E_{max}, \textbf{x}, \Omega)$, where $E_{max}$ is some maximal initial energy. 

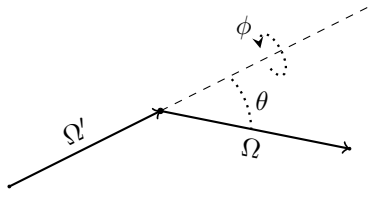
\begin{figure}[h]
	\centering
	\begin{tikzpicture}
		\filldraw[shift={(0, 0), color=black}] circle (0.5 pt) node (A) {} node[anchor=north]{};
		\filldraw[shift={(2, 1), color=black}] circle (1 pt) node (B) {} node[anchor=west]{};
		\filldraw[shift={(4.5, 0.5), color=black}] circle (0.5 pt) node (C) {} node[anchor=west]{};
		\filldraw[shift={(5, 2.5), color=white}] circle (0.0 pt) node (D) {} node[anchor=west]{};
		
		\draw [thick, ->=.5] (A.center) -- (B.center) node [midway, above, sloped] (line1) {$\Omega'$};
		\draw [dashed] (B.center) -- (D.center) node [midway, sloped] (line2) {\AxisRotator[dotted]};
		\draw [thick, ->=.5] (B.center) -- (C.center) node [midway, below, sloped] (line3) {$\Omega$};
		\draw [dotted, line width=.2ex] (3.2, 0.75) arc (0:45:1) node [midway, right] (line1) {$\theta$};
		\draw (3.1,2.1) node {$\phi$};
	\end{tikzpicture}
	\caption{Angular deflections at scattering events \cite{nuclear_energy_agency_penelope_2019}.}
	\label{fig:RotationalDeflections}
\end{figure}

The quantity $\Sigma_t (E, \textbf{x})$, referred to as the total scattering cross-section, is measured in terms of collisions per distance unit. In other words, on average, particles undergo $\Sigma_t (E, \textbf{x})$ collisions per distance unit they travel through the medium. When particles undergo a collision, their new velocity is given by the screened Rutherford elastic Differential Cross-Section (DCS) $\Sigma_s(E, \textbf{x}, \Omega \cdot \Omega')$ from \cite{kawrakow_egsnrc_nodate} with \begin{align}
	\label{eq:DCS}
	\Sigma_t (E, \textbf{x}) = \int \Sigma_s(E, \textbf{x}, \mu) d\mu,
\end{align} where $\mu = \Omega \cdot \Omega' = \cos{\theta}$ is the cosine of the polar scattering angle $\theta$. More specifically, when a particle collides with the background tissue, its orientation after the collision is given by a polar scattering angle and a uniformly distributed azimuthal scattering angle $\phi$ (see figure \ref{fig:RotationalDeflections}). To make this quantitative, an orientation $\Omega'$ is expressed in spherical coordinates \begin{align}
	\Omega' = \begin{bmatrix}
		\sin(\upvartheta) \cos(\upvarphi) \\
		\sin(\upvartheta) \sin(\upvarphi) \\
		\cos(\upvartheta)
	\end{bmatrix} = \begin{bmatrix}
		u' \\
		v' \\
		w'
	\end{bmatrix},
\end{align} where the angles $\upvarphi$ and $\upvartheta$ are the angles with the z and x-axis respectively \cite{nuclear_energy_agency_penelope_2019}. The quantities $u$, $v$, and $w$ are the velocities in the x, y, and z directions. After a collision, the particle's new orientation is \begin{align}
	\label{eq:RotateVelocity}
	\Omega = R(\upvarphi \mathbf{\hat{z}}) R(\upvartheta \mathbf{\hat{y}}) \begin{bmatrix}
		\sin(\theta) \cos(\phi) \\
		\sin(\theta) \sin(\phi) \\
		\cos(\theta)
	\end{bmatrix},
\end{align} where $R(\upvarphi \mathbf{\hat{z}})$ and $R(\upvartheta \mathbf{\hat{y}})$ are matrices which apply a rotation about the z and y axis respectively.  \\

For completeness, we note that the collision integral in \eqref{eq:BoltzmannCSD} can be rewritten in standard form \cite{jablonski_monte_2005} in which the scattering rate and post-collisional orientation distribution become visible \begin{multline}
	\int \Sigma_s(E, \mathbf{x}, \Omega \cdot \Omega') \psi(E, \textbf{x}, \Omega') d\Omega' \\
	= \int \Sigma_s(E, \mathbf{x}, \cos(\theta)) \frac{\Sigma_t(E, \mathbf{x})}{\Sigma_t(E, \mathbf{x})} \psi(E, \textbf{x}, \Omega) 2 \pi \sin(\theta) d\theta \end{multline} \begin{align}
	\qquad \qquad \qquad \qquad \; \; = \int \underbrace{\frac{2 \pi \sin(\theta) \Sigma_s(E, \mathbf{x}, \cos(\theta))}{\Sigma_t(E, \mathbf{x})}}_{\text{post-collisional velocity distribution}} \Sigma_t(E, \mathbf{x}) \psi(E, \textbf{x}, \Omega) d\theta \label{eq:postcolRadiation} .
\end{align} The first factor in the integral in equation \eqref{eq:postcolRadiation} is an explicit expression of the (normalized) velocity distribution of a particle after undergoing a collision.  

The kinetic behavior of the particles lends itself nicely to Monte Carlo simulation. A particle's initial position and orientation are sampled from the initial condition $\psi(E_{max}, \textbf{x}, \Omega)$. Particles then travel along straight lines to their next collision point. The distance between two collision points is sampled from an exponential distribution with rate $\Sigma_t (E, \textbf{x})$. Upon a scattering collision, a polar and azimuthal scattering angle are sampled, after which the new orientation can be computed as a rotation of the previous velocity \eqref{eq:RotateVelocity}. The energy loss due to the step can be obtained by integrating the stopping power over the path \cite{nuclear_energy_agency_penelope_2019}. This process is repeated until the particle reaches some small threshold energy level, at which the particle is removed from the simulation. As the particle travels through space, the dose (deposited energy) is scored at each collision. The stochastic process that is described here, is called a velocity-jump process \cite{othmer_diffusion_2000}. Algorithms that simulate the velocity-jump process explicitly are referred to as analog particle tracing algorithms. This simulation algorithm is quite simple and easily parallelizable. However, because every collision is executed explicitly, the algorithm becomes computationally infeasible when the total scattering cross-section $\Sigma_t (E, \textbf{x})$ is large. That is the motivation for kinetic-diffusion-type algorithms, which are introduced in section \ref{section:KD}. For an overview of other particle tracing algorithms, see \cite{mortier_advanced_2020}.

\subsection{Comparison to linear Boltzmann equation with independent velocities}
\label{section:fusionComparison}

In this section, we present the linear Boltzmann equation \cite{cercignani_boltzmann_1988} as it is used in e.g. fusion energy and highlight the differences with the CSD Boltzmann equation \eqref{eq:BoltzmannCSD} used in electron beam therapy \cite{cercignani_boltzmann_1988}. Understanding the differences and similarities between the models will ultimately reveal how the Kinetic-Diffusion Monte Carlo algorithm can be adapted to the kinetic equation for electron beam therapy.

The density of a single species of neutral particles in the plasma edge of a nuclear fusion reactor $\varphi_n(\textbf{x}, \textbf{v}, t)$ is governed by the following linear Boltzmann equation \cite{mortier_advanced_2020} \begin{multline}
	\label{eq:BoltzmannNeutrals}
	\underbrace{\pdv{\varphi_n(\textbf{x},\textbf{v},t)}{t}}_{\text{transient term}} + \underbrace{ \vphantom{\pdv{\varphi_n(\textbf{x},\textbf{v},t)}{t}} \textbf{v} \grad_{\textbf{x}} \varphi_n(\textbf{x},\textbf{v},t)}_{\text{transport}} = -\underbrace{ \vphantom{\pdv{\varphi_n(\textbf{x},\textbf{v},t)}{t}} R_i(\textbf{x},\textbf{v})\varphi_n(\textbf{x},\textbf{v},t)}_{\text{sink due to ionisation}} - \underbrace{ \vphantom{\pdv{\varphi_n(\textbf{x},\textbf{v},t)}{t}} R_{cx}(\textbf{x},\textbf{v})\varphi_n(\textbf{x},\textbf{v},t)}_{\text{sink due to scattering}} \\ + \underbrace{f_{\text{postcol}}(\textbf{v}|\textbf{x}) \int R_{cx}(\textbf{x},\textbf{v}')\phi (\textbf{x},\textbf{v}',t)d\textbf{v}'}_{\text{source due to scattering}} ,
\end{multline} with position $\mathbf{x} \in \mathbb{R}^3$, velocity $\mathbf{v} \in \mathbb{R}^3$ and time $t \in \mathbb{R}_{\geq 0}$. Note that this velocity variable $v$ is not normalized, as was the case for the orientation $\Omega$ in equation \eqref{eq:BoltzmannCSD}. The left-hand side of equation \eqref{eq:BoltzmannNeutrals} describes the advection through space. The time variable is conceptually the same as the energy variable in electron beam therapy; the main difference being that the energy decreases with a rate given by the stopping power $S(E, \textbf{x})$, whereas the time variable increases with rate one. The right-hand side of equation \eqref{eq:BoltzmannNeutrals} contains a source term and two sinks. The sink due to ionization removes particles from the neutral population at rate $R_i(\textbf{x},\textbf{v})$, and thus models an absorption collision. The remaining sink and source together model a scattering collision analogous to equation \eqref{eq:BoltzmannCSD}. As in electron beam therapy, when the collision rate $R_{cx}(\textbf{x},\textbf{v})$ becomes large, Monte Carlo simulation of the particle process becomes computationally infeasible. 

The most important difference between the linear Boltzmann equation and the CSD Boltzmann equations lies in the post-collisional velocity distribution. Comparing the terms of the collision integrals \eqref{eq:postcolRadiation} and \eqref{eq:BoltzmannNeutrals}, we note that in the case of the CSD Boltzmann equation subsequent velocities are correlated, i.e. the post-collisional velocity distribution depends on the pre-collisional velocity through the polar scattering angle. In the case of equation \eqref{eq:BoltzmannNeutrals}, subsequent velocities are independent: the post-collisional velocity distribution $f_{\text{postcol}}(\textbf{v}|\textbf{x})$ can be put outside of the collision integral in \eqref{eq:BoltzmannNeutrals}. 

In conclusion, the linear Boltzmann equation is similar to the kinetic equation used in the domain of electron beam therapy. However, the difference in how the post-collisional velocities are sampled will require special attention when porting the kinetic-diffusion Monte Carlo algorithm to the field of electron beam therapy. This process, along with the KDMC algorithm itself, is described in section \ref{section:KD}.

\theoremstyle{remark}
\newtheorem{remark}{Remark}

\section{Kinetic-Diffusion-type Algorithms}
\label{section:KD}

\subsection{Kinetic-Diffusion Monte Carlo for linear Boltzmann equation with independent velocities}
\label{section:KDMC}
Consider a particle in a homogeneous medium, following the dynamics dictated by the kinetic equation \eqref{eq:BoltzmannNeutrals}, i.e., particles undergo collisions with rate $R_{cx}(\textbf{x},\textbf{v})$, and upon a collision, particles obtain a new velocity sampled from $f_{\text{postcol}}(\textbf{v}|\textbf{x})$. Ionization collisions are ignored. In a high collisional regime, the kinetic-diffusion Monte Carlo algorithm avoids computing every collision explicitly by aggregating multiple collisions into one step. In mathematical terms, we try to find an approximation for the distance $\mathbf{\Delta x} \in \mathbb{R}^3$ a particle travels in a fixed time step $\Delta t \in \mathbb{R}^{+}$ \begin{align}
	\label{eq:deltaX}
	\mathbf{\Delta x} = \sum_{j=0}^{J} \Delta t_j \mathbf{v}_j \; \; \; \text{with}\; \; \;  \Delta t = \sum_{j=0}^{J} \Delta t_j,
\end{align} where $\Delta t_j$ is the time between two collisions, $\mathbf{v}_j \in \mathbb{R}^3$ is the velocity of the particle after the $j$-th collision and $J$ is the number of collisions a particle undergoes within a fixed time $\Delta t$. KDMC is based on the observation that when the number of collisions $J$ in equation \eqref{eq:deltaX} becomes large, due to the central limit theorem \cite{blitzstein_introduction_2019}, the (stochastic) positional increment $\mathbf{\Delta x}$ can be expected to become normally distributed. Thus, in the limit, the following is an approximation for $\mathbf{\Delta x}$: \begin{align}
	\label{eq:basicRandomWalk}
	\mathbf{\Delta x} \approx \bm{\mu}(\Delta t) + \bm{\sigma}(\Delta t) \bm{\xi},
\end{align} with $\bm{\mu}(\Delta t) \in \mathbb{R}^3$ and $\bm{\sigma}(\Delta t) \in \mathbb{R}^{3 \times 3}$ the time-step dependent mean and standard deviation of the approximate normal distribution to $\mathbf{\Delta x}$, and $\bm{\xi} \in \mathbb{R}^3$ a standard normally distributed number. Note that this is equivalent to tracing the particle using only the first two moments of the true stochastic process. For the remainder of this text, a positional increment as in equation \eqref{eq:basicRandomWalk} is referred to as a \emph{biased random walk}. 

The argument above was formalized in \cite{mortier_kinetic-diffusion_2022}. There, using a Hilbert expansion of the particle distribution, it was found that as the scattering rate tends to infinity, the linear Boltzmann equation \eqref{eq:BoltzmannNeutrals} behaves according to an advection-diffusion equation. This limit model is often referred to as a diffusion limit. Since the particle dynamics associated with an advection-diffusion equation are of the form \eqref{eq:basicRandomWalk}, its use is justified \cite{mortier_advanced_2020}. Expressions for the mean $\bm{\mu}(\Delta t)$ and variance $\bm{\sigma}^2(\Delta t)$ were then obtained by computing the mean and variance of the velocity-jump process for a fixed, finite collision rate $R_{cx}(\textbf{x},\textbf{v})$.

The principle of kinetic-diffusion Monte Carlo is to use so-called kinetic-diffusion (KD) steps. These steps hybridize explicit simulation of the velocity-jump process (kinetic steps), which simulate the Boltzmann equation \eqref{eq:BoltzmannNeutrals}, with the random walk steps, which simulate the diffusion limit. Specifically, the hybridization occurs as follows. Particle tracks are divided into time steps of size $\Delta t$. In each time step, the particle first executes a kinetic step \begin{align}
	\mathbf{x}_j' = \mathbf{x}_j + \Delta t_j \mathbf{v}_j.
\end{align}
Then, if the collision occurs before the end of the time step, the particle moves with a random walk step for the remainder of $\Delta t$ \begin{align}
	\mathbf{x}_{j+1} = \mathbf{x}_j' + \max(\Delta t - \Delta t_j, 0)\left[\bm{\mu}(\Delta t - \Delta t_j) + \bm{\sigma}(\Delta t - \Delta t_j) \bm{\xi} \right].
\end{align}
Note that every KD step consists of one kinetic step until the first collision, and at most one random walk step for the remainder of the time step. The computational cost thus becomes independent of the collision rate. Note that for infinitely small collision rates, the time-step $\Delta t_j$ will always be larger than $\Delta t$, such that the complete step is performed with the kinetic scheme. Similarly, in the case of infinitely large collision rates, the kinetic step is infinitely small and the only contribution to the motion of the particle is the biased random walk step. In this limit, the diffusion limit is valid. Algorithms that are exact in both asymptotically low and high collisional regimes, are referred to as asymptotic preserving. The downside is that the random walk step introduces a modelling error that scales with $\mathcal{O}\left(\sqrt{R_{cx}^3 \Delta t^3}\right)$. The modelling error and convergence of the scheme was analysed in \cite{mortier_kinetic-diffusion_2022}.

\subsection{Kinetic-Diffusion Monte Carlo for electron beam therapy}
\label{section:KDMCRT}
In this section, we port KDMC for the linear Boltzmann equation to the Boltzmann CSD equation for electron beam therapy. In section \ref{section:meanAndVarianceRadiation}, we derive the mean and variance of the random walk step for a velocity-jump process with dependent velocities. In section \ref{section:MSdistribution}, we introduce the concept of a multiple scattering distribution. In section \ref{section:KDR}, the full Kinetic-Diffusion-type algorithm is presented. We refer to the algorithm as Kinetic-Diffusion-Rotation due to the rotational dependencies between subsequent velocities. Finally, in section \ref{section:KDRGeneralizations}, we discuss how the KDR algorithm deals with boundary conditions, heterogeneous media, and more complex particle dynamics.~

\subsubsection{Mean and variance of the kinetic motion}
Consider a particle in a homogeneous medium, following the dynamics dictated by the kinetic equation \eqref{eq:BoltzmannCSD}. The particle is initially oriented along the direction $\Omega_{-1}$ and scatters with scattering rate $\Sigma_t$. Let the average polar deflection angle upon a collision be denoted as $\expect{\cos(\theta)}$. We write the net change in the position of the particle, under the constraint that the total traveled distance $\Delta s$ is fixed, as \begin{align}
	\mathbf{\Delta x} = \sum_{j=0}^{J} \Delta s_j \Omega_j \; \; \; \text{with}\; \; \;  \Delta s = \sum_{j=0}^{J} \Delta s_j, 
\end{align} where the $\Delta s_j$ are the stepsizes between two collisions points and $\Omega_j$ the orientations of the particles. Note the similarity with section \ref{section:KDMC}, $\Delta s$ takes on the role of the fixed time $\Delta t$, and total scattering cross-section $\Sigma_t$ takes on the role of the scattering rate $R_{cx}$. In what follows, we give an expression for the mean motion $\bm{\mu}(\Delta s)$ and discuss a suitable approximation to the variance $\bm{\sigma}(\Delta s)$, see \eqref{eq:basicRandomWalk}.

\begin{figure}[h]
	\begin{subfigure}[b]{0.49\textwidth}
		\centering
		\includegraphics[width=\linewidth]{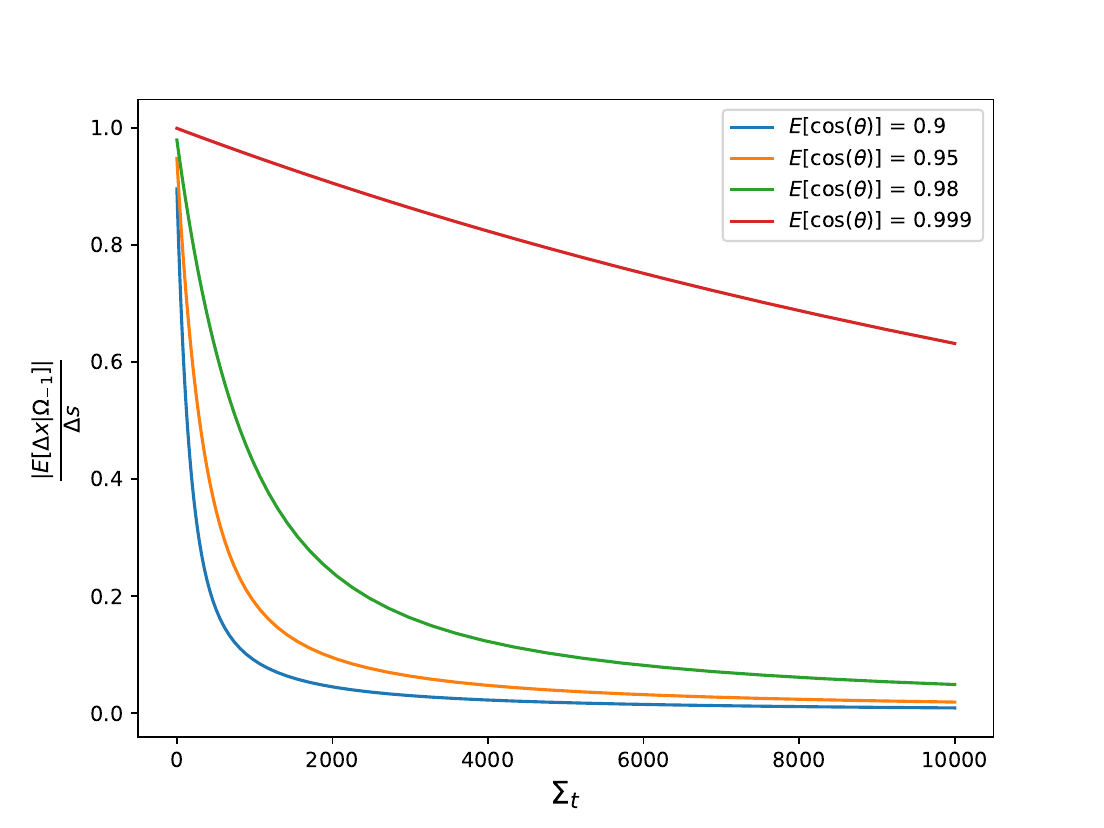}
		\subcaption{Scaled norm of the mean of kinetic motion as a function of the scattering rate for various values of $\expect{\cos(\theta)}$.}
		\label{fig:theoreticalMeanKineticMotion}
	\end{subfigure}
	\begin{subfigure}[b]{0.49\textwidth}
		\centering
		\includegraphics[width=\linewidth]{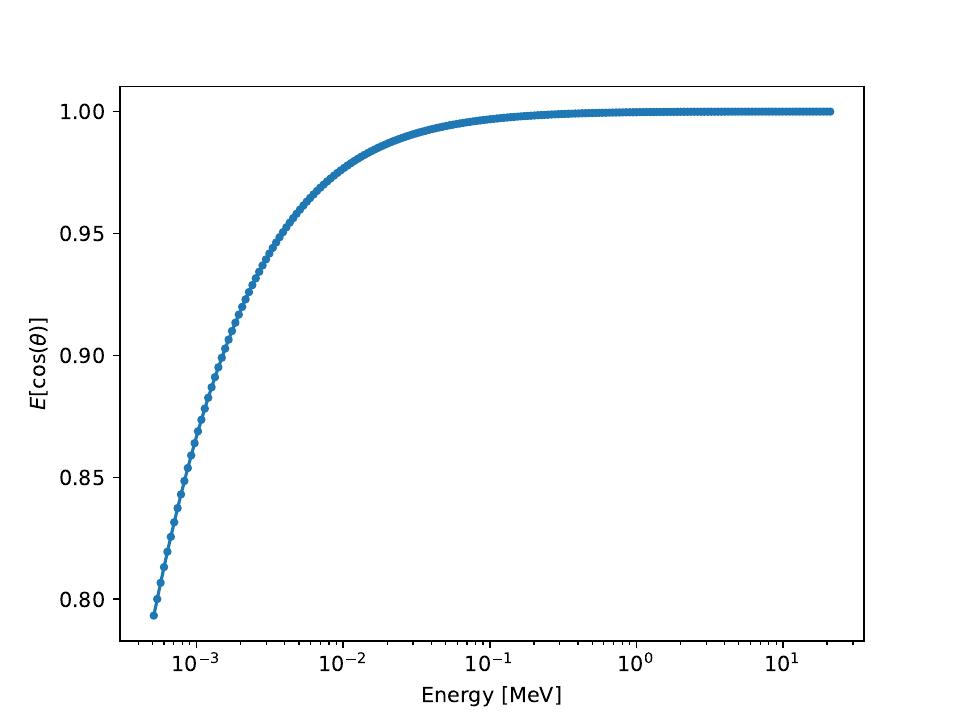}
		\subcaption{Expectation of the cosine of the polar scattering angle of the screened Rutherford scattering cross-section.}
		\label{fig:Ecost}
	\end{subfigure}
	\caption{Mean kinetic motion (left) and the dependence of the mean polar scattering angle on the energy (right). }
\end{figure}

\label{section:meanAndVarianceRadiation}
\textbf{Mean} \newline
The average positional increment is then given by \begin{align}
	\label{eq:KDRMean}
	\expect{\Delta x | \Omega_{-1}} = \Omega_{-1} \frac{\expect{\cos(\theta)}}{1 - \expect{\cos(\theta)}} \frac{1}{\Sigma_t}  \left[ 1 - e^{\Sigma_t \Delta s( \expect{\cos(\theta)} - 1)} \right] , 
\end{align} The proof for \eqref{eq:KDRMean} is given in appendix \ref{section:meanRadiationProof} and assumes no energy loss throughout the time step (fixed energy). Thus, the total scattering cross-section \(\Sigma_t\) and probability density function for \(\cos(\theta)\), given by the differential scattering cross section \(\Sigma_s\) \eqref{eq:DCS}, are fixed throughout a time step. As one would expect due to the rotational symmetry, the mean of the kinetic motion lies along the initial direction $\Omega_{-1}$. The distance a particle travels in the direction $\Omega_{-1}$ is determined by the mean of the polar scattering angle distribution $\expect{\cos(\theta)}$, the scattering rate $\Sigma_t$ and step size $\Delta s$. The proportion of the mean $\expect{\Delta x | \Omega_{-1}}$ with respect to $\Delta s$ is plotted as a function of the scattering rate in figure \ref{fig:theoreticalMeanKineticMotion} for several values of $\expect{\cos(\theta)}$. Larger scattering rates yield less advection. This is due to the physical interpretation that if a particle undergoes many random deviations, the particle is expected to travel far less in the expected direction. Note that the above formulation heavily depends on \(\Sigma_t\) being homogeneous in space. Generalisation to non-homogeneous media is discussed in section \ref{section:KDRGeneralizations}.

\begin{figure}[h]
	\centering
	\includegraphics[width=0.6\linewidth]{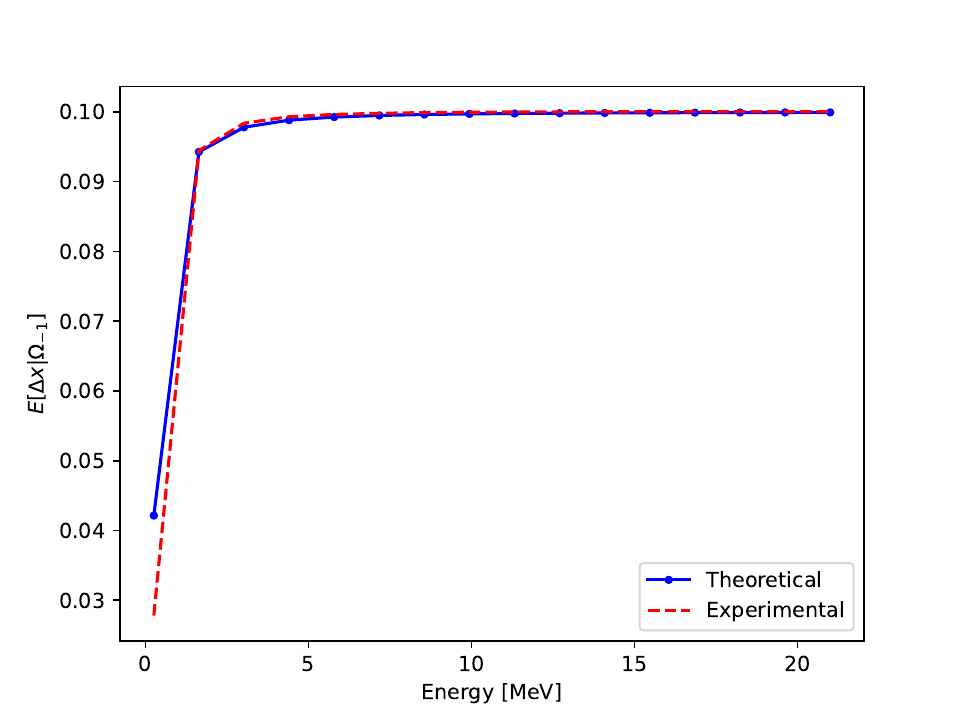}
	\caption{Comparison of experimentally obtained mean kinetic motion (10.000 particles) with the theoretical result \eqref{eq:KDRMean} as a function of the energy. Particles are set to travel a distance $\Delta s = 0.1$.}
	\label{fig:MeanKineticMotionValidation}
\end{figure}

We validate the formula for the mean of the kinetic motion against a kinetic simulation with 10.000 particles. The particles are emitted from a point source at the origin along the z-axis in a homogeneous medium made of water. When particles have covered a distance of $\Delta s= 0.1$, their position is recorded. Due to the symmetry, the average motion of the particle is the z-coordinate of its final position. This experimentally obtained value is compared to the theoretical formula \eqref{eq:KDRMean}. The results are plotted in figure \ref{fig:MeanKineticMotionValidation}. For large energy values, the theoretical formula fits the data well, confirming the correctness of the result. For small values of the energy, the theoretical result overestimates the mean movement. This is because the theoretical result assumes no energy loss of the particle during the random walk step. For small values of the energy, the expectation of the cosine of the polar scattering angle $\expect{\cos(\theta)}$ is strongly dependent on the energy (see figure \ref{fig:Ecost}), resulting in an error on the mean motion.

\newpage
\textbf{Variance} \newline
Unfortunately, the procedure to derive the mean position increment does not readily generalize to the variance. In short, the proof for the variance requires the expectation value $\expect{\Omega_j^2 | \Omega_{-1}}$. This expectation value requires the knowledge of the distribution of the orientation after $j$ collisions. To the author's knowledge, there does not exist an analytic formula for this distribution, nor does there exist a useful formula for the expectation. An incomplete derivation for the variance is provided in \cite{willems_particle_2023}. 

Instead of using an analytic formula, we store the variance of the kinetic motion in a lookup table. The lookup table is generated as follows. A point source at the origin emits particles along the z-axis. Particles are simulated collision by collision until they have covered a distance $\Delta s$, upon which the final position of the particle is recorded. In a post-processing step, the variance in the x, y, and z-direction on the final position is computed and stored in the lookup table. This process is repeated for a linearly spaced range of initial energy values $E_{max}$, step sizes $\Delta s$, and background media with density $\rho$. The background medium is characterized solely by its density; the chemical composition is that of water and is constant throughout the domain. The specific values for which the lookup table was generated are summarized in table \ref{tab:VarLUT}. We note that in the case of electron beam radiation therapy, the energies of the electrons and the densities of the background media are known in advance. Therefore, the lookup tables are not problem-dependent and can be reused. 

\begin{figure}[h!]
	\centering
	\begin{subfigure}[b]{0.30\textwidth}
		\centering
		\captionsetup{justification=centering}
		\includegraphics[width=\linewidth]{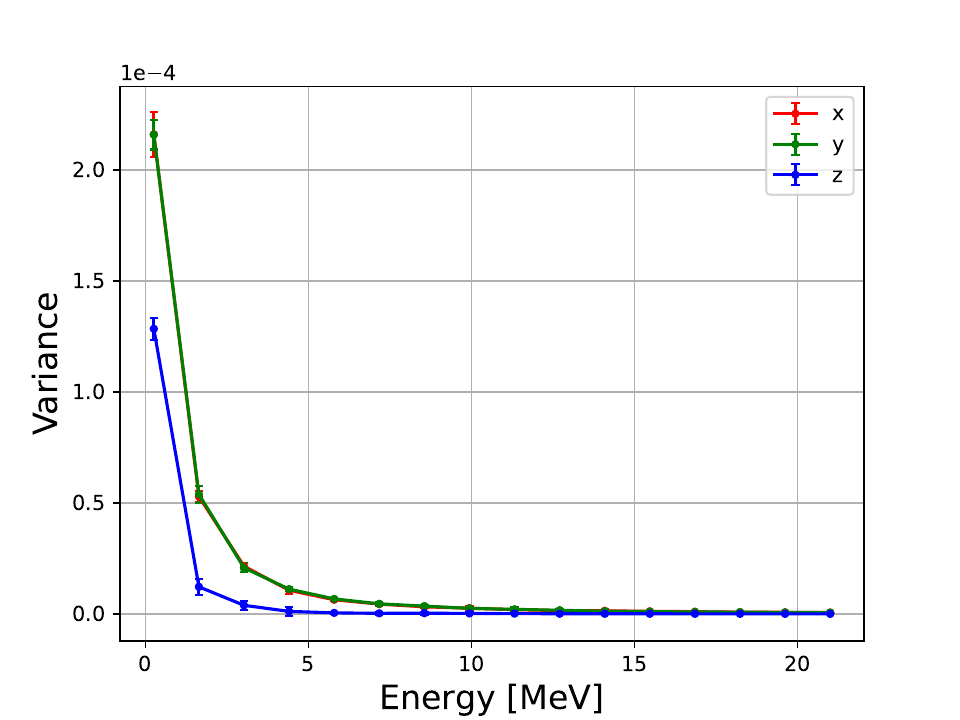}
		\subcaption{$\Delta s = 5.4\times10^{-2} \; cm$, $\rho = 1.01 \; g/cm^3$.}
		\label{fig:VarLUTe}
	\end{subfigure}
	\begin{subfigure}[b]{0.30\textwidth}
		\centering
		\captionsetup{justification=centering}
		\includegraphics[width=\linewidth]{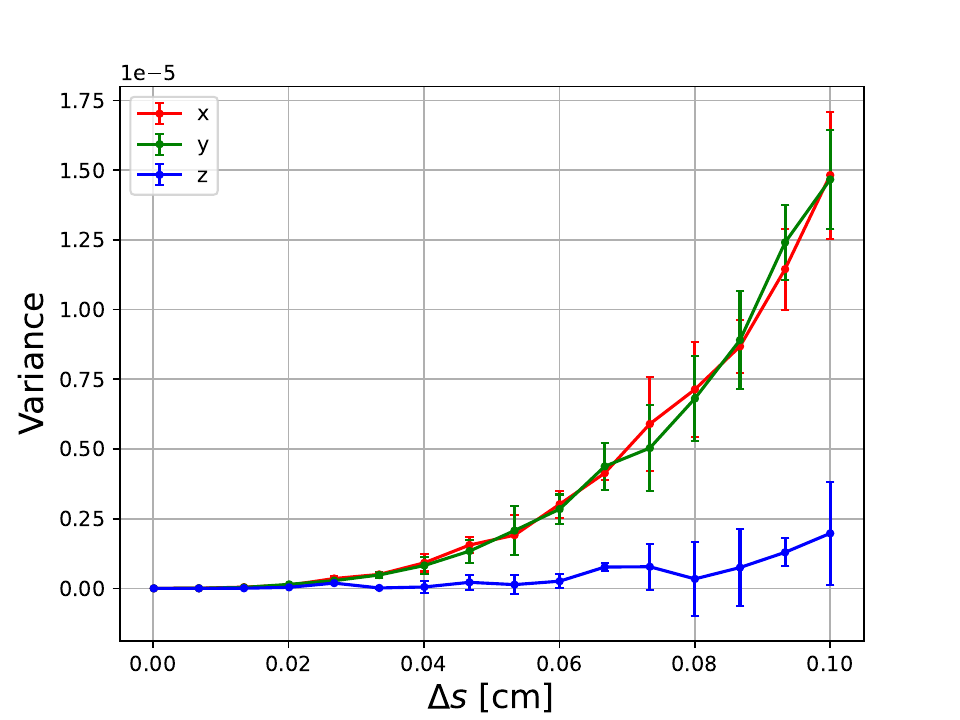}
		\subcaption{$E = 11.32$ MeV, \\$\rho = 1.01 \; g/cm^3$.}
		\label{fig:VarLUTds}
	\end{subfigure}
	\begin{subfigure}[b]{0.30\textwidth}
		\centering
		\captionsetup{justification=centering}
		\includegraphics[width=\linewidth]{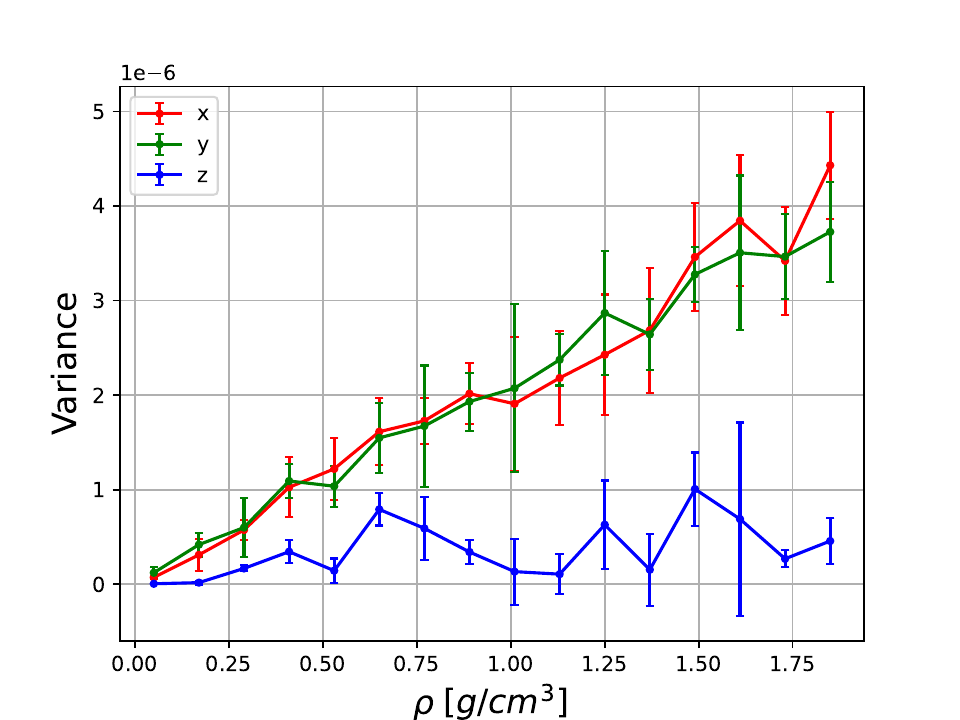}
		\subcaption{$E = 11.32$ MeV,  \\$\Delta s = 5.338 \times 10^{-2}$.}
		\label{fig:VarLUTrho}
	\end{subfigure}
	\caption{Lookup table for the variance in the x, y and z direction generated with a kinetic particle tracing algorithm. Specific simulation parameters are summarized in table \ref{tab:VarLUT}. The error bars represent one standard deviation computed using the batch means method with 10 batches \cite{flegal_batch_2010}.}
	\label{fig:VarLUT}
\end{figure}

The variance is plotted for several values of the energy, step size, and density in figure \ref{fig:VarLUT}. For a fixed step size and background medium, a larger initial energy yields particles with little deviation from their original orientation (lower variance). Larger step sizes lead to more accumulated collisions, and thus a larger variance. Due to symmetry, the variance in the x and y direction is the same. The variance in the z-direction contains more statistical noise than in the x and y directions. When the variance is required in a direction $\Omega_{-1}$ that is not the z-axis, the variance is rotated from the z-axis to $\Omega_{-1}$. This is explained in detail in section \ref{section:KDR}.

\begin{table}[h]
	\centering
	\begin{tabular}{ |c|c|c|c|c|c| } 
		\hline
		$E_{min}$ & $E_{max}$ & $\rho_{min}$ & $\rho_{max}$ & $\Delta s_{min}$ & $\Delta s_{max}$\\ 
		\hline
		0.5 MeV & 21 MeV & $0.05 \; g/cm^3$ & $1.85 \; g/cm^3$ & $1\times10^{-4} \; cm$ & $1 \; cm$ \\ 
		\hline
	\end{tabular}
	\begin{tabular}{ |c|c|c|c|c|c| } 
		\hline
		Amount of bins for $E$, $\rho$ and $\Delta s$ & Amount of particles \\ 
		\hline
		16 & $10^4$ \\ 
		\hline
	\end{tabular}
	\caption{Parameters used for the generation of the lookup table for the variance of the kinetic motion. }
	\label{tab:VarLUT}
\end{table}

\subsubsection{Multiple scattering distribution}
\label{section:MSdistribution}
The kinetic-diffusion-rotation algorithm, which is discussed in the next section, achieves improved performance when a multiple scattering (MS) distribution is considered. An MS distribution is the distribution of the orientation after a particle has covered a distance $\Delta s$ in which it possibly had multiple scattering collisions, given some initial orientation $\Omega_{-1}$. In general, this distribution is unknown because it is a solution to the Boltzmann equation itself. There exist many theories on suitable approximations of the multiple scattering distribution, such as the Fokker-Planck approximation, Molière theory and the Goudsmit-Saunderson distribution \cite{vassiliev_monte_2017}. 

\begin{figure}[h!]
	\centering
	\begin{subfigure}[b]{0.49\textwidth}
		\centering
		\includegraphics[width=\linewidth]{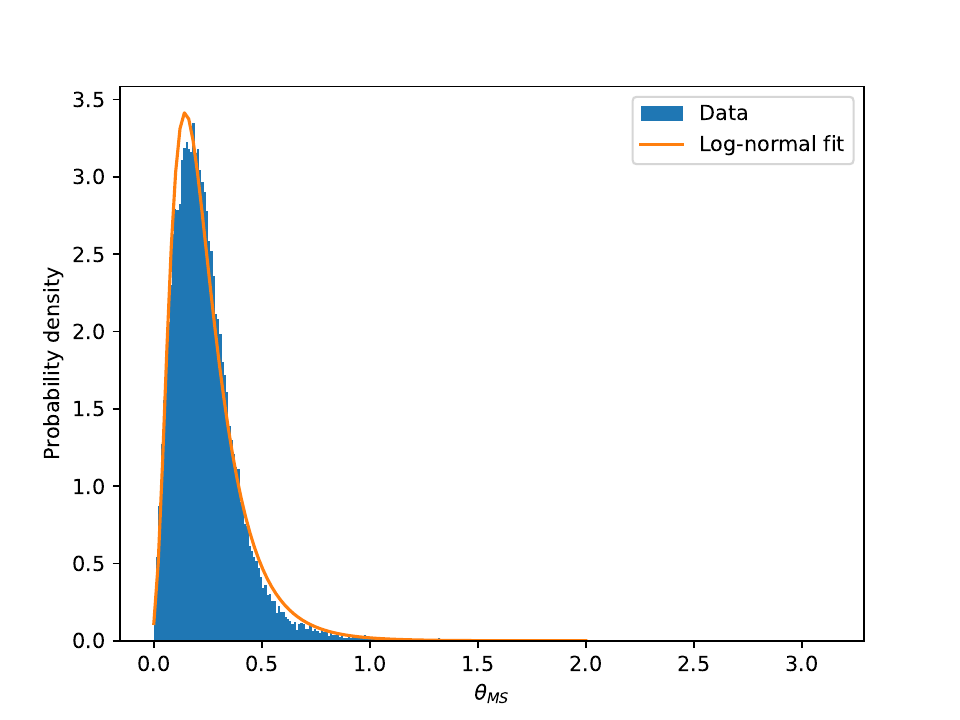}
		\subcaption{Distribution and fit.}
		\label{fig:thetaMS}
	\end{subfigure}
	\begin{subfigure}[b]{0.49\textwidth}
		\centering
		\includegraphics[width=\linewidth]{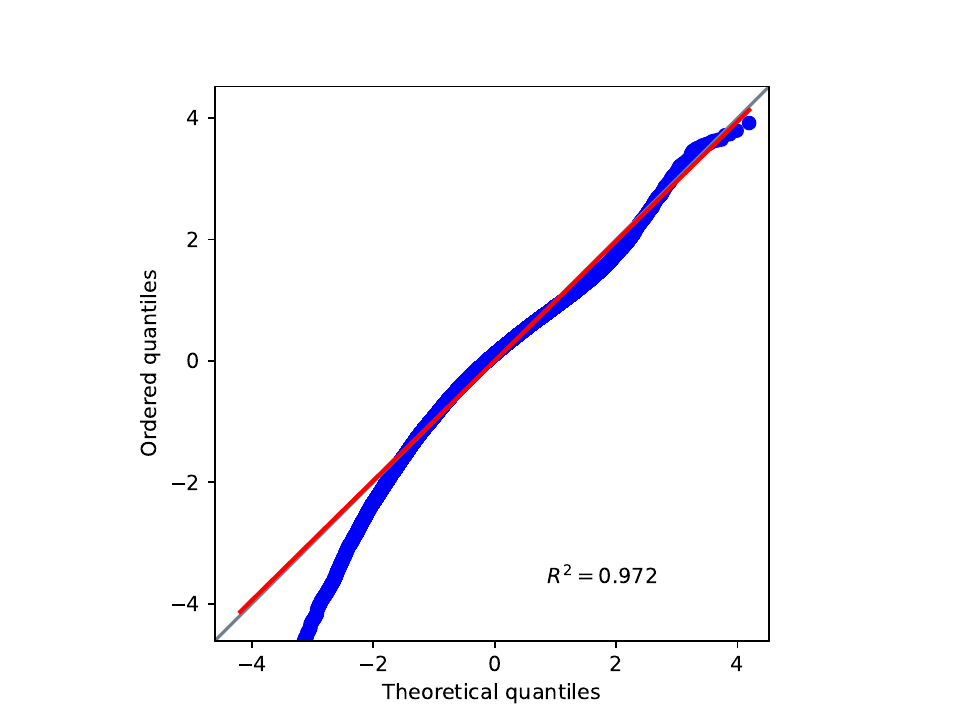}
		\subcaption{Q-Q plot with log-normal distribution.}
		\label{fig:lognormalQQ}
	\end{subfigure}
	\caption{Polar multiple scattering distribution $\theta_{MS}$ for a particle with energy $E_{max} = 3.117$ MeV, that has travelled $\Delta s = 0.1$ centimetres in a medium with density $\rho = 1.0 \; g/cm^3$. The distribution is fitted with a log-normal distribution.}
	\label{fig:LogMS}
\end{figure}

For simplicity, we obtain the multiple scattering distribution in a data-driven approach. Similar to the variance, the multiple-scattering distribution is approximated by tracing 50.000 particles for a distance $\Delta s$. Particles are spawned at a point source at the origin with energy $E_{max}$ and are oriented along the z-axis. Using a kinetic particle tracing algorithm, particles are simulated until they have covered a distance $\Delta s$. At that point, their orientation is stored. This process is repeated for different step sizes $\Delta s$, energies $E_{max}$, and background media $\rho$. Since the azimuthal scattering angle is uniformly distributed, symmetry around the z-axis is expected. The distribution of the orientation is therefore expressed in terms of the polar multiple scattering angle $\theta_{MS}$. This is the angle between the final orientation of a particle and its initial orientation (z-axis). The angle $\theta_{MS}$ and a log-normal fit are plotted in figure \ref{fig:thetaMS}. In figure \ref{fig:lognormalQQ}, a Q-Q plot \cite{wilk_probability_1968} of $\theta_{MS}$ is plotted. The Q-Q plot indicates that the polar multiple scattering angle distribution is approximated well by the log-normal distribution.

\subsubsection{Kinetic-Diffusion-Rotation Monte Carlo}
\label{section:KDR}
Where KDMC performs steps with fixed time $\Delta t$, KDR performs steps with a fixed traveled distance $\Delta s$. Each KDR step has two components: a kinetic step and a random walk step. First, a kinetic step is executed. Then, if a collision occurs before the end of the step, the remainder of the step is executed with a random walk. If no collision occurred, two options exist. The first would be to truncate the kinetic step $\Delta s$. This would not entail a bias since the step lengths $\Delta s_k$ are exponentially distributed, which leads to the memoryless property \cite{nuclear_energy_agency_penelope_2019}. The second option would be to execute the kinetic step completely. The subsequent diffusive step can then be taken of size $\Delta s - (\Delta s_k \;  \; \text{MOD} \; \; \Delta s)$. The latter strategy is preferred since it requires fewer random numbers to be sampled. Due to this combination of kinetic and random walk step, the majority of the motion in a low-collisional regime is kinetic. In a high-collisional regime, the random walk dominates the behavior of the particle. In both regimes, the execution time of KDR is constant, since always one kinetic step and one random walk step are performed. 

The random walk step is characterized by an advection and diffusion coefficient, for which we choose the mean and variance of the kinetic motion as discussed in section \ref{section:meanAndVarianceRadiation}. We denote the distance the particle travels using the random walk step as $d_k$.  Because the lookup table for the variance is generated for a particle aligned with the z-axis, the variance must be rotated in the direction of the kinetic step $\Omega_{k}$. The rotation matrix $R(\phi_{k} \mathbf{\hat{z}}) R(\theta_{k} \mathbf{\hat{y}})$ \eqref{eq:RotateVelocity}, with $\theta_{k}$ and $\phi_{k}$ the polar and azimuthal angle of $\Omega_{k}$, achieves this effect.  A KDR step in three dimensions is then characterized by the following equations, \begin{align}
	\mathbf{x}_k' &= \mathbf{x}_k + \Delta s_k \Omega_k, \\
	\mathbf{x}_{k+1} &= \mathbf{x}_k' + A_k d_k + R(\phi_{k} \mathbf{\hat{z}}) R(\theta_{k} \mathbf{\hat{y}}) \sqrt{V} \mathbf{\xi} ,
\end{align} with \begin{align}
	d_k &= \Delta s - (\Delta s_k \;  \; \text{MOD} \; \; \Delta s) \label{eq:},\\
	\Delta s_k &\sim \mathcal{E}(\Sigma_t) ,\\
	V &= diag(\sigma_x^2, \sigma_y^2, \sigma_z^2) ,\\
	\mathbf{\xi} &\sim \mathcal{N}(\mathbf{0}, I_3) ,\\
	A_k &= \frac{\Omega_{k} \expect{\cos(\theta)}}{1 - \expect{\cos(\theta)}}\frac{1}{\Sigma_t t_k} \left[ 1 - e^{\Sigma_{t} t_k( \expect{\cos(\theta)} - 1)} \right] ,\label{eq:KDMCAdvectionCoef}
\end{align} where $I_3$ is the three-by-three unit matrix, $\mathcal{E}$ the exponential distribution and $\mathcal{N}$ the normal distribution. The equations above yield the position of the particle after the KDR step. The energy after the KDR step can be computed using existing techniques for the kinetic step and the random walk step at once. For simplicity, we stick to an approximation based on Euler's rule \cite{kawrakow_egsnrc_nodate, hissoiny_gpumcd_2011} that produces an $\mathcal{O}(\Delta E^2)$ error, \begin{align}
	\label{eq:computeEnergy}
	E_k - E_{k+1} = S \left( E_k - S(E_k) \frac{\Delta s}{2} \right) \Delta s ,
\end{align} where $E_k$ is the energy at the beginning of the KDR step and $S$ is the stopping power from equation \ref{eq:BoltzmannCSD}. Next, the particle tracing routine needs to return the orientation of the particle after the KDR step $\Omega_{k+1}$. Two possible solutions are considered. Firstly, the orientation of the particle after the random walk step can be approximated by applying a (single-scattering) rotation based on the vector that connects $x_k'$ and $x_{k+1}$. Alternatively, the orientation $\Omega_{k+1}$ can be sampled from a multiple scattering distribution. Both solutions are tested in section \ref{section:numerical}. The full KDR algorithm is summarised in appendix \ref{section:kdrAlg}. 

\subsubsection{Generalizations of KDR to a practical algorithm}
\label{section:KDRGeneralizations}
Up to this point, we have only considered a simple scattering particle in an infinite homogeneous medium. Generalizations from this simplified model are discussed here. Firstly, in a practical setting, domains are finite and typically have periodic or reflective boundaries. Boundary hits during the kinetic step of KDR are dealt with in the usual way \cite{mortier_advanced_2020}. However, boundary hits during the random walk step, require extra care. In case the random walk step (line 15 in algorithm \ref{alg:KDR}) exceeds the domain boundaries, it is not executed. Instead, the particle is traced to the boundary kinetically. 

In addition to having boundary conditions, practical domains are heterogeneous. Specifically, in the case of electron beam therapy, domain information is provided using CT scans. These CT scans define a three-dimensional rectangular grid in which material parameters such as density and chemical composition are piecewise constant \cite{du_plessis_indirect_1998}. When a particle crosses from one grid cell to another during a kinetic step, the particle is paused at the grid cell boundary such that a new step size can be sampled using the new material parameters. This is allowed due to the Markovian property of velocity-jump processes \cite{nuclear_energy_agency_penelope_2019}. However, when a grid cell crossing occurs during a random walk step, the particle cannot be paused on the boundary since the position increment is not given by an exponentially distributed step. In addition, rejecting a random walk step outside the grid cell would entail a bias. There is no choice but to execute the random walk step. This strategy can cause errors near material boundaries since the part of the motion in the neighboring grid cell does not take into account the new background. To avoid numerical artifacts, in all numerical experiments the step size $\Delta s$ is taken smaller than the grid size.

Finally, it is common for particles to undergo multiple types of collisions such as scattering, absorption, and emission. The algorithm as presented above does not yet consider multiple types of events. In case another type of event occurs before the end of the random walk step, the step can be replaced by kinetic motion. An alternative strategy is to only execute the random walk step to the collision point \cite{mortier_advanced_2020}.

\section{Numerical results with KDR}
\label{section:numerical}
In this section, a prototype implementation of KDR \cite{willems_electrontransportcode_2023} is tested on two benchmark problems. All tests are performed using the particle model described in section \ref{section:radiationModel}, with one simplification: The scattering rate and polar angle distribution are fixed as if the particle has an energy of 2.61 MeV. This assumption, although unphysical, reduces the computation time required for generating the lookup tables of the multiple scattering distribution and the variance. 

In section \ref{section:KDRSpeedUp}, the speed-up of KDR compared to an analog particle tracing algorithm is discussed. Finally, in section \ref{section:KDRLung}, the KDR algorithm is applied to a dose estimation problem in electron beam therapy. In addition, this test illustrates the impact of using a multiple scattering distribution.

\subsection{Empirical speed-up}
\label{section:KDRSpeedUp}
\begin{figure}[h]
	\centering
	\includegraphics[width=0.5\linewidth]{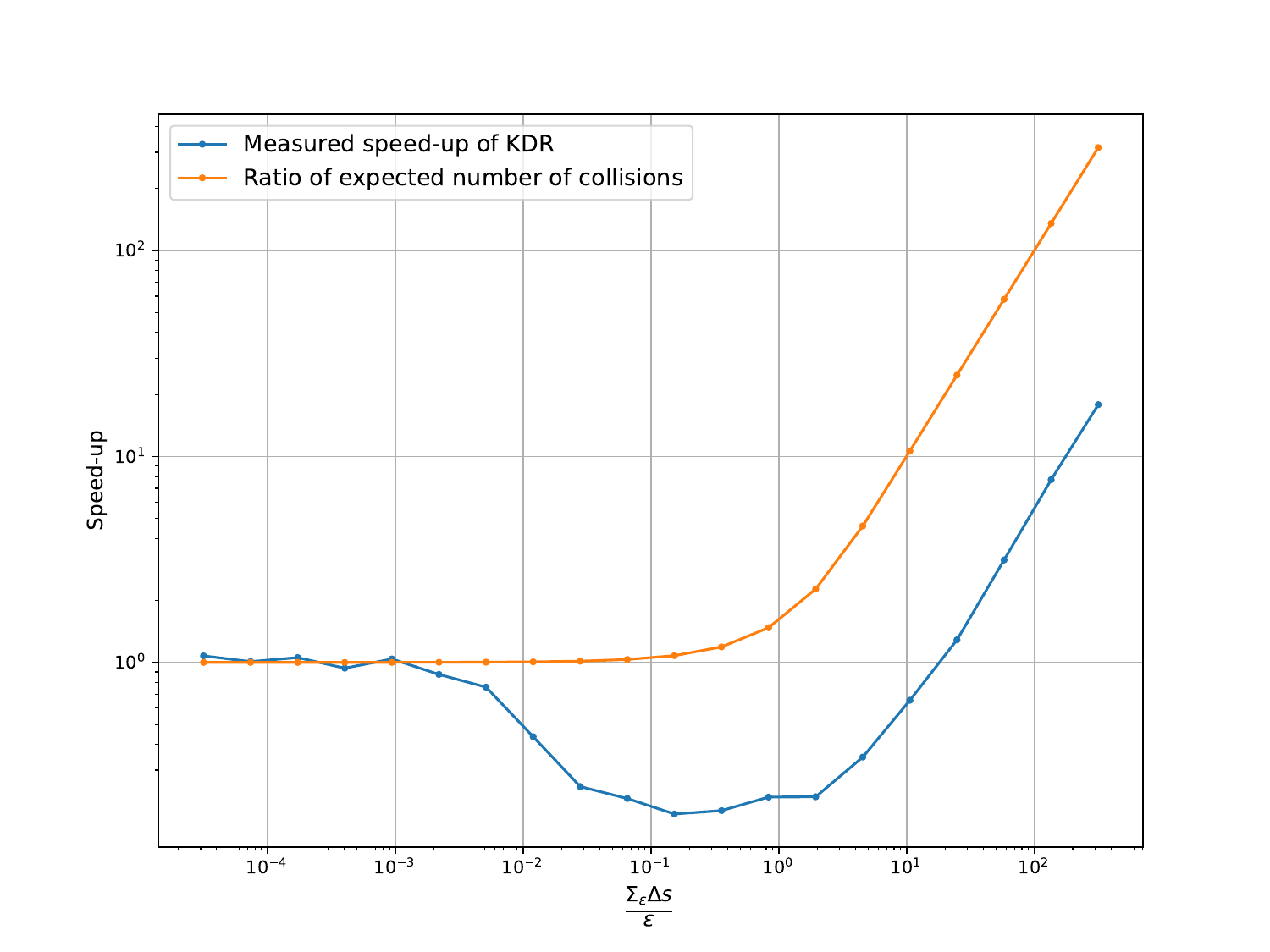}
	\caption{Speed-up of KDR compared to purely kinetic simulation for a particle with scattering characteristics of a 2.61 MeV electron.}
	\label{fig:KDRSpeedUp}
\end{figure}

KDR achieves a speed-up by aggregating multiple kinetic steps into one diffusive step. A theoretical speed-up can be derived based on the expected number of collisions in a simulation. In a kinetic simulation, the expected number of collisions within a step $\Delta s$ is $\Sigma_t \Delta s$. In a KD simulation, only the first collision is computed explicitly. In that case, the expected number of collisions equals the probability of at least one collision occurring in a kinetic simulation, which equals $1 - e^{-\Sigma_t \Delta s}$ \cite{mortier_advanced_2020}. The ratio of the expected number of collisions is a theoretical speed-up. In figure \ref{fig:KDRSpeedUp}, the theoretical speed-up is compared to empirically obtained speed-up. 

For small values of the collisionality $\Sigma_t \Delta s$, KDR performs similarly to the analog particle tracing algorithm. This is because the simulation is completed in one kinetic step, so no diffusive steps are performed. As the scattering rate increases, the diffusive step is executed more often and the speed-up temporarily decreases. This is because, for a relatively small step, a diffusive step is more computationally expensive than a kinetic step. For large scattering rates, the diffusive step aggregates many kinetic steps and then becomes computationally efficient.  

\subsection{Practical use case}
\label{section:KDRLung}
\begin{table}[h]
	\centering
	\begin{tabular}{ |c|c|c|c|c|c|c|c| } 
		\hline
		$E_{max}$ & $\sigma_E$ & $d$ & $\sigma_z$ & $\kappa_\theta$ & $\Delta s$ & $N$\\ 
		\hline
		21 MeV & $\frac{1}{100}$ & 14.5 $cm$ & $\frac{1}{50}$ & 10000 & 0.0725 & 100.000 \\ 
		\hline
	\end{tabular}
	\caption{Simulation parameters for the lung test case with KDR. The initial energy of the particle is given by a normal distribution with mean $E_{max}$ and standard deviation $\sigma_E$. The domain has the shape of a square with size $d$ in the y-z direction and is infinite in the x direction. The initial position of the particle is given by (0, $d$, $z_i$), where $z_i$ is sampled from a Gaussian distribution with mean $d/2$ and standard deviation $\sigma_z$. For each simulation algorithm, $N$ particles are simulated. The KDR stepsize is $\Delta s$. The initial velocity of the particles is constrained to the y-z plane where the angle with the negative y-axis is given by a von Mises distribution with center zero and dispersion $\kappa_\theta$.}
	\label{tab:KDRLung}
\end{table}
As a proof-of-concept, we examine the application of the kinetic-diffusion-rotation algorithm to dose estimation in the domain of electron beam therapy. To this end, a 2D CT scan of a lung patient is radiated by an electron beam \cite{kusch_robust_2021}. The 2D CT scan of the lung is expanded in the third dimension to allow full 3D particle tracing. The initial position and energy of the particles are given by \begin{align}
	\psi(E, \mathbf{x}) = \frac{1}{2\pi \sigma_{E} \sigma_z} e^{-\frac{1}{2}\left(\frac{E - E_{max}}{\sigma_E}\right)^2} \delta(x) \delta(y - d) e^{-\frac{1}{2} \left( \frac{z + d/2}{\sigma_z} \right)^2 } .
\end{align} The initial velocity of the particles is constrained to the y-z plane where the angle with the negative y-axis is given by a von Mises distribution with center zero and dispersion $\kappa_\theta$. As previously mentioned, the particle's scattering characteristics are fixed as if it were a particle at 2.61 MeV. The background medium is assigned the chemical composition of water and the density of the medium is derived from the pixel values of the CT scan \cite{kupper_models_2016, kusch_robust_2021}. A white pixel (255) is assigned the density of bone $\rho_{bone} = 1.85 g/cm^3$. A completely black pixel (0) is assigned a minimum density of $\rho_{min} = \; 0.05 \; g/cm^3$. The density of the other pixels is linearly scaled based on the pixel value. All relevant simulation parameters are repeated in table \ref{tab:KDRLung}. The dose is estimated using the analog particle tracing algorithm, KDR with multiple scattering distribution, and KDR without multiple scattering distribution. All simulations are performed on the Flemisch supercomputer using sixteen MPI processes. 
\begin{figure}[h!]
	\centering
	\begin{subfigure}[b]{0.49\textwidth}
		\centering
		\includegraphics[width=\linewidth]{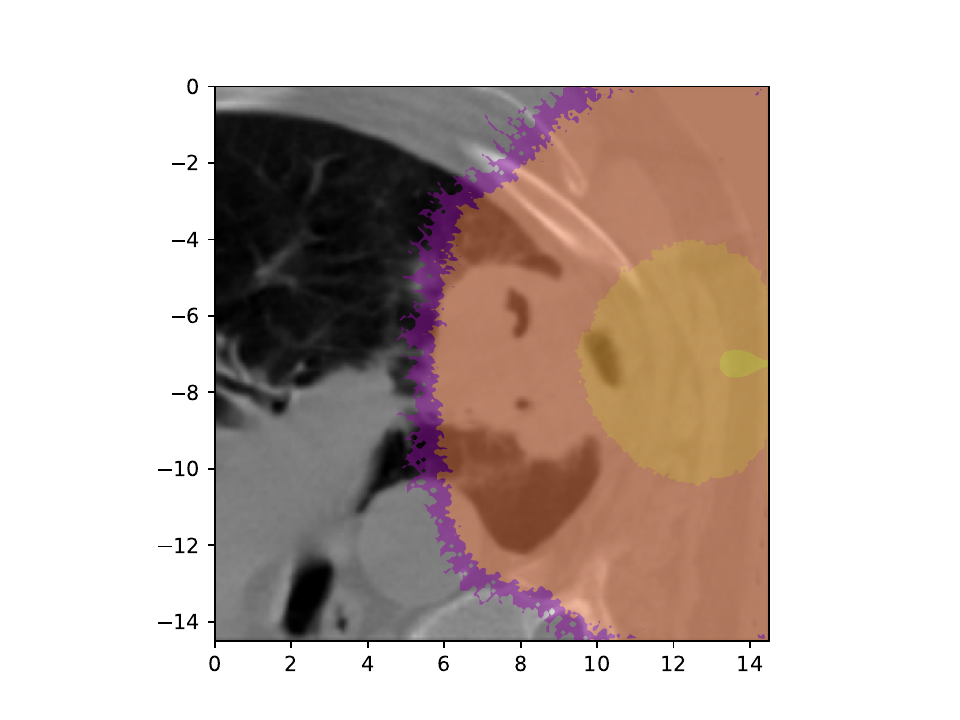}
		\subcaption{Analog simulation.\newline}
		\label{fig:KDRLungK}
	\end{subfigure}
	\begin{subfigure}[b]{0.49\textwidth}
		\centering
		\includegraphics[width=\linewidth]{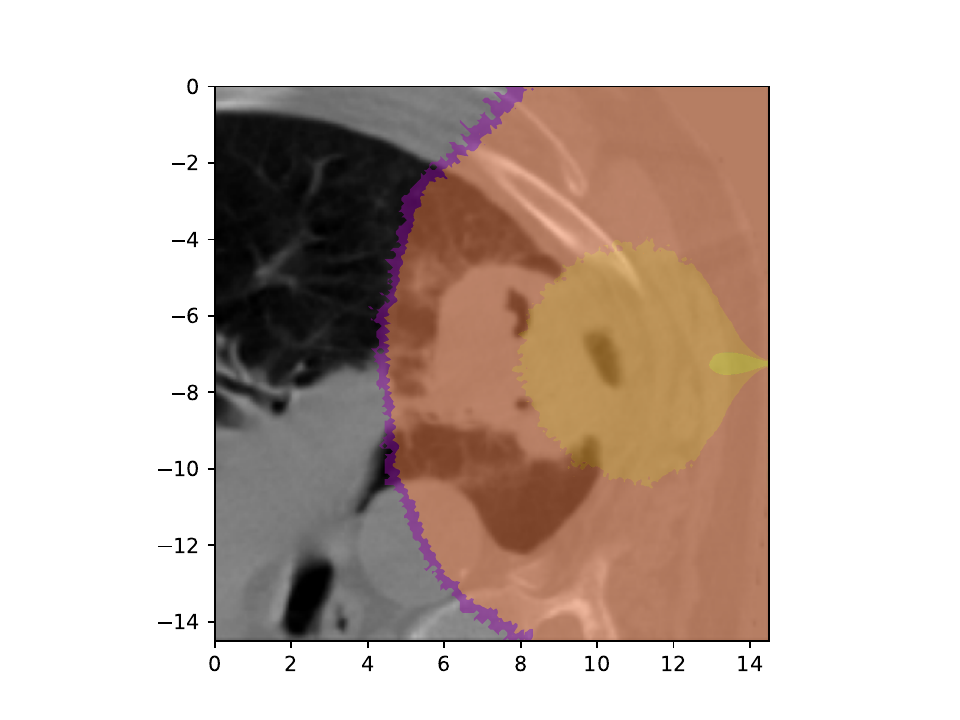}
		\subcaption{KDR without multiple scattering distribution.}
		\label{fig:KDRLungKDR}
	\end{subfigure}
	\begin{subfigure}[b]{0.49\textwidth}
		\centering
		\includegraphics[width=\linewidth]{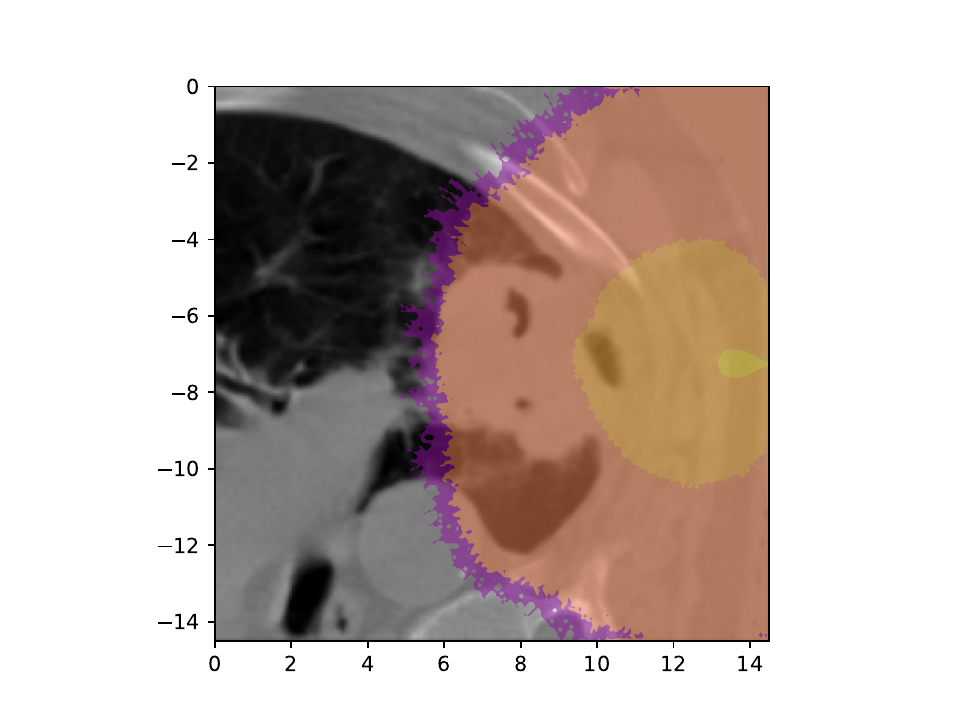}
		\subcaption{KDR with multiple scattering distribution.}
		\label{fig:KDRLungKDRMS}
	\end{subfigure}
	\caption{2D CT scan of a lung and the computed dose distribution using three different particle tracing algorithms. The borders of the colored regions are isodose lines for $10^{-8}, 10^0, 10^2, 10^3$, and $10^5 \frac{\text{MeV}}{\text{cm}^2}$. All relevant simulation parameters are summarised in table \ref{tab:KDRLung}. }
	\label{fig:KDRLung}
\end{figure}

The dose distribution is projected onto the y-z plane and plotted in figure \ref{fig:KDRLung}. The borders of the colored regions are isodose lines for $10^{-8}, 10^0, 10^2, 10^3$, and $10^5 \frac{\text{MeV}}{\text{cm}^2}$. The dose distribution for KDR without multiple scattering distribution does not match the result obtained with the analog particle distribution. The borders of the colored regions are isodose lines. The isodose lines at the edges are close together, indicating a steep drop in the dose distribution. In the center the isodose lines are far apart, indicating a relatively flat dose distribution. In addition, the area of the dose distribution is noticeably larger than for the analog particle tracing algorithm, indicating too much advection. These issues are resolved when a multiple scattering distribution is used. 

\begin{figure}[h!]
	\centering
	\begin{subfigure}[b]{0.49\textwidth}
		\centering
		\includegraphics[width=\linewidth]{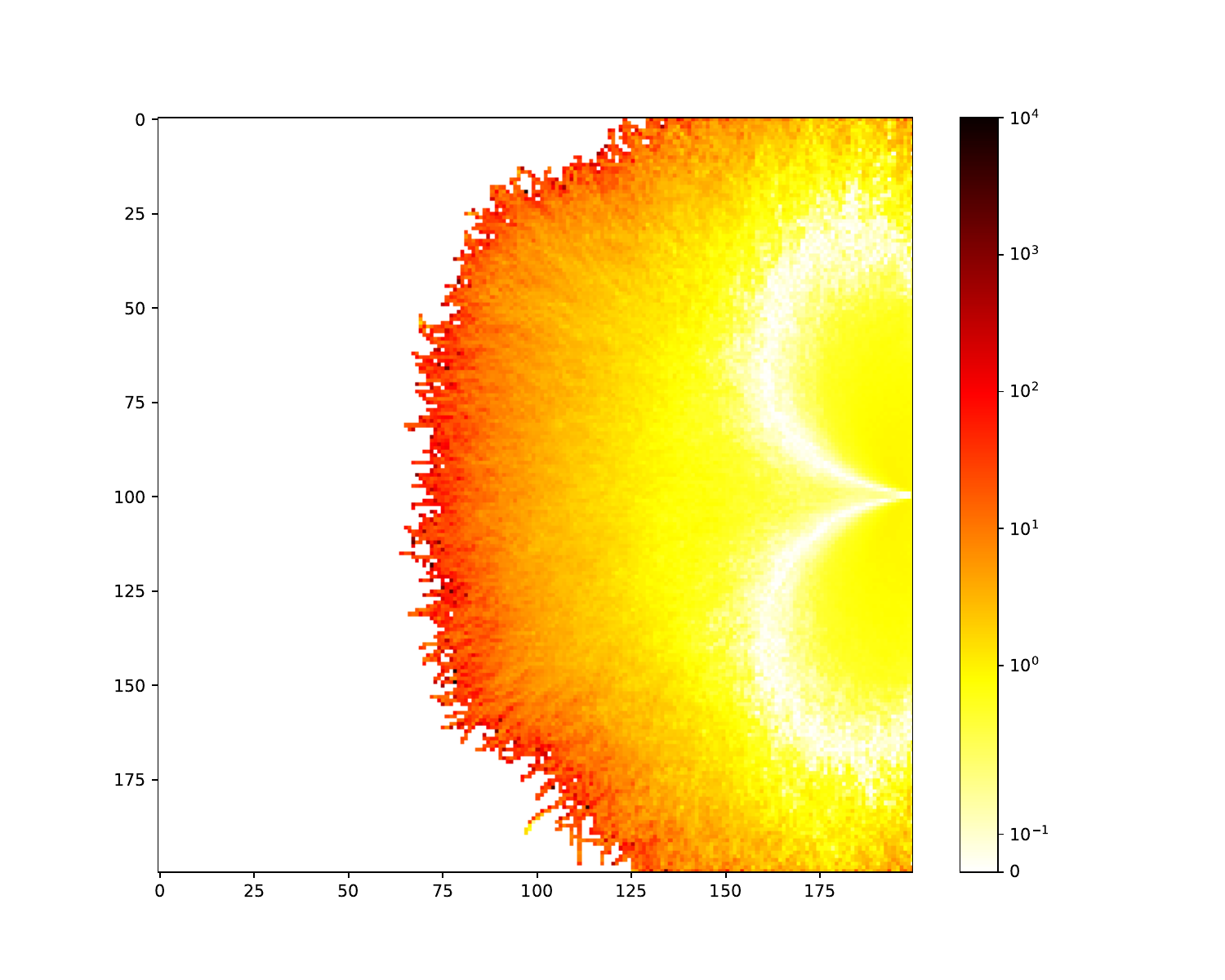}
		\subcaption{KDR without multiple scattering distribution.}
		\label{fig:LungKDRError}
	\end{subfigure}
	\begin{subfigure}[b]{0.49\textwidth}
		\centering
		\includegraphics[width=\linewidth]{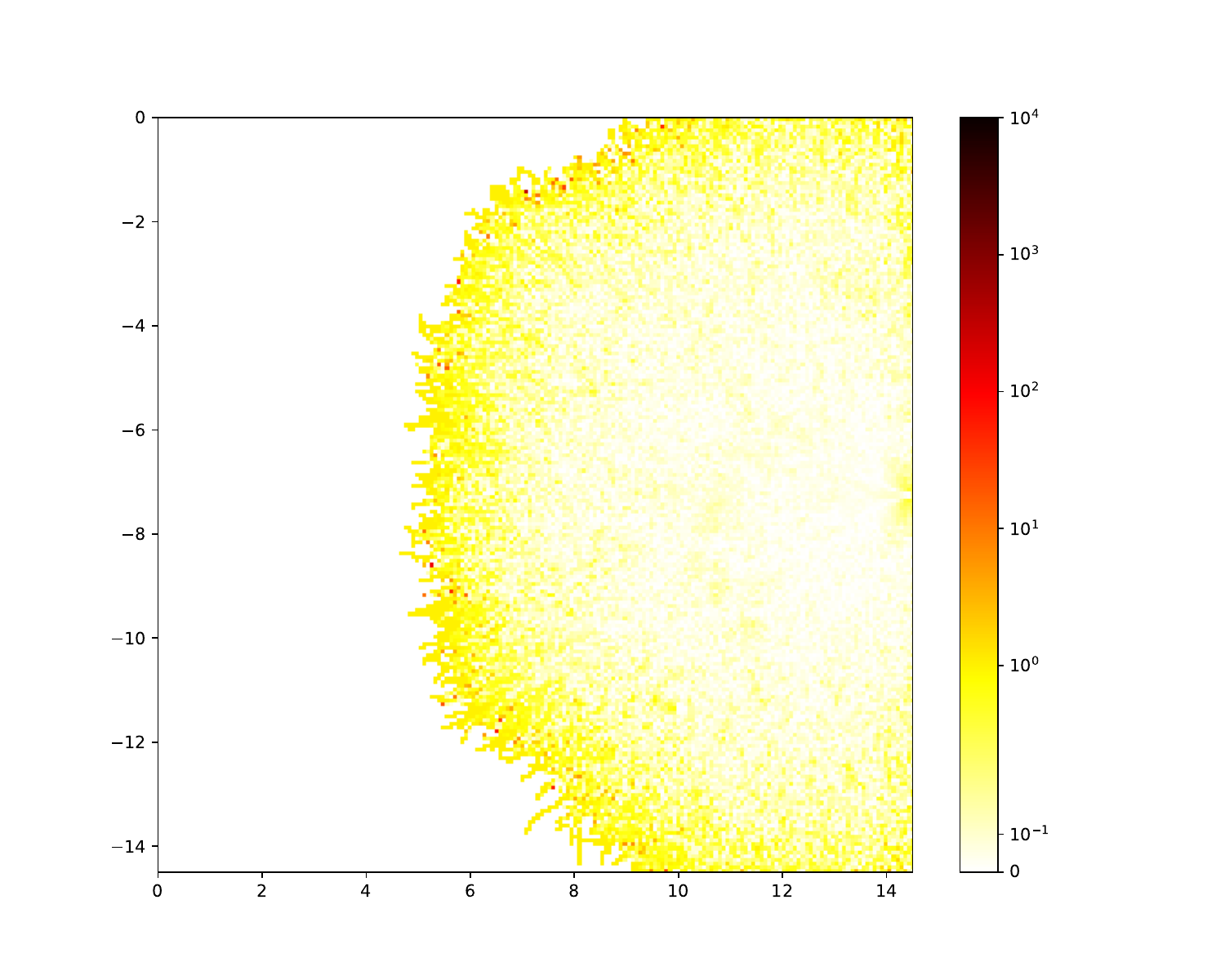}
		\subcaption{KDR with multiple scattering distribution.}
		\label{fig:LungKDRMSError}
	\end{subfigure}

	\caption{Pointwise relative error of the dose distributions.}
	\label{fig:KDRLungError}
\end{figure}

In figure \ref{fig:KDRLungError}, the pointwise relative error of the dose distributions with respect to the analog simulation is plotted in a heatmap. For both algorithms, the region near the edge of the radiated tissue contains the largest errors. This is because the particles are unlikely to reach this region, thus, the statistical error is largest there. However, note that in the case of KDR without multiple scattering distribution the errors are significantly higher. The average relative error of the dose distributions on the whole domain for KDR and KDR MS are 60\% and 7.7\% respectively. 
\begin{table}[h]
	\centering
	\begin{tabular}{ |c|c|c| } 
		\hline
		Analog & KDR & KDR MS \\ 
		\hline
		06:12:51 & 00:09:33 & 00:11:20 \\ 
		\hline
	\end{tabular}
	\caption{Timings results for the lung test case obtained on the Flemisch supercomputer using 16 MPI processes. Results are given in the hh:mm:ss format.}
	\label{tab:KDRLungtimings}
\end{table}

Timings results for the 2D lung test case are summarized in table \ref{tab:KDRLungtimings}. KDR with MS distribution is approximately 19\% slower than KDR without MS distribution. This is because the parameters of the multiple scattering distribution must be fetched from a lookup table before a new velocity must be sampled. Note, however, that KDR with multiple scattering distribution still achieves a speed-up of nearly 33 compared to the analog particle tracing algorithm.

\clearpage
\section{Conclusion}
\label{section:conclusion}
Monte Carlo methods are the state-of-the-art for dosimetric computations in electron beam therapy. However, their execution time becomes a bottleneck in highly collisional regimes. To address this, we introduced a kinetic-diffusion particle tracing scheme. Originally proposed for the linear Boltzmann equation in the context of neutral transport for fusion energy, this algorithm explicitly simulates kinetic motion in low-collisional regimes and dynamically switches to a random walk in high-collisional regimes. The random walk is constructed to preserve the first two moments (mean and variance) of the kinetic motion. We derived an analytic expression for the mean kinetic motion and considered incorporating a multiple scattering distribution into the scheme. Unlike in neutral transport, in the radiative transfer setting an analytical expression for the variance is not easily obtained, so we instead used a lookup table. We tested the algorithm for dosimetric calculations on a 2D CT scan of a lung patient. With a simple particle model, our Python implementation achieved a speedup of nearly 33× over a purely kinetic simulation, with a relative modeling error of 7.7\%.For this work, we used a simplified particle model featuring only a single type of scattering event. Extending the kinetic-diffusion approach to more detailed particle models—such as those in PENELOPE \cite{nuclear_energy_agency_penelope_2019} and EGS \cite{w_r_egs4_1990} — remains an avenue for future research.

\textbf{Acknowledgement.} The resources and services used in this work were provided by the VSC (Flemish Supercomputer Center), funded by the Research Foundation - Flanders (FWO) and the Flemish Government. The first author has received funding from the European Union's Framework Program for Research and Innovation HORIZON-MSCA-2021-DN-01 under the Marie Sklodowska-Curie Grant Agreement Project 101072546 – DATAHYKING. The second author is an SB PhD fellow of the Research Foundation Flanders (FWO), funded by grant 1S64723N. The authors would like to thank Prof. Jonas Kusch for providing us with the \href{https://github.com/JonasKu/publication-A-robust-collision-source-method-for-rank-adaptive-dynamical-low-rankapproximation}{test case} used in section \ref{section:KDRLung}.
\clearpage

\begin{appendices}
\section{Proof for the mean motion in KDR }
\label{section:meanRadiationProof}
In this section, the mean of the kinetic motion $\expect{\Delta x | \Omega_{-1}}$ is derived, given some initial orientation of the particle $\Omega_{-1}$. Compared to \cite{mortier_estimation_2022}, we take into account the rotational dependency between subsequent velocities. This mean can then be used in the KDR algorithm as a replacement for the advection coefficient \eqref{eq:basicRandomWalk}. The derivation is presented in three dimensions. Throughout the derivation, it is assumed that the scattering rate $\Sigma_t$ and polar scattering angle $\mu = \cos(\theta)$ are constant within the diffusive step. 

We take the same approach as in \cite{mortier_kinetic-diffusion_2022, willems_particle_2023} but exchange the time variable $t$ from fusion for the pathlength variable $s$. The length of the trajectory of a particle is divided into steps of fixed size $\Delta s$. During an interval $[k \Delta s, (k+1) \Delta s]$, particles undergo $J$ collisions at distances $s_j$ from the start of the interval. The distance between two collisions is denoted by $\Delta s_j$. The amount of collisions $J$ is Poisson distributed with average $\Delta s \Sigma_s$. Thus, analogous to the case in fusion, we obtain for each interval the following equations \begin{align}
	\Delta x &= \sum_{j=0}^{J} \Delta s_j \Omega_j,  \\
	\Delta s &= \sum_{j=0}^{J} \Delta s_j,  \\
	P(J = j) &= \frac{\left( \Sigma_t \Delta s\right)^j}{j!} e^{-\Sigma_t \Delta s}.
\end{align}
Since in radiation therapy, each velocity depends on all previous velocities (see section \ref{section:radiationModel}), the mean is conditioned on the velocity of the particle $\Omega_{-1}$ before the interval. In the context of KDMC, where kinetic and diffusive steps are alternated, the velocity $\Omega_{-1}$ is the velocity of the previous kinetic step. Using the law of total expectation, the mean motion of the particle conditioned on the previous velocity can be written as \begin{align}
	\expect{\Delta x | \Omega_{-1}} &= \expect*{ \sum_{j=0}^{J} \Delta s_j \Omega_j | \Omega_{-1}} \label{eq:KDRMeanStart}\\
	&= \expect^J*{ \expect*{\sum_{j=0}^{J} \Delta s_j \Omega_j | J, \Omega_{-1} } | \Omega_{-1}} \\
	&= \sum_{J=0}^{\infty} \frac{\left( \Sigma_t \Delta s \right)^J}{J!} e^{-\Sigma_t \Delta s} \expect*{\sum_{j=0}^{J} \Delta s_j \Omega_j | J, \Omega_{-1} } \label{eq:KDRmeantemp2}.
\end{align} Using the linearity property of the expectation value, the remaining expectation value in \eqref{eq:KDRmeantemp2} simplifies to \begin{align}
	\label{eq:KDRmeantemp1}
	\expect*{\sum_{j=0}^{J} \Delta s_j \Omega_j | J, \Omega_{-1} } &= \sum_{j=0}^{J} \expect{\Delta s_j | J} \expect{\Omega_j | \Omega_{-1}} .
\end{align} The first factor in \eqref{eq:KDRmeantemp1} is the mean step size, given the total number of collisions $J$ and was derived in \cite{mortier_advanced_2020}: \begin{align}
	\expect{\Delta s_j | J} = \frac{\Delta s}{J+1} \label{eq:KDMC_dsk_exp}.
\end{align} The second factor in \eqref{eq:KDRmeantemp1} is the mean of the $j$-th velocity $\Omega_j$, given the previous velocity $\Omega_{-1}$. Before this expectation can be solved, consider the more simple expectation value $\expect{\Omega_0 | \Omega_{-1}}$. The distribution of velocity $\Omega_0$ conditioned velocity $\Omega_{-1}$ is given by the matrix equation \eqref{eq:RotateVelocity}. Written out in the components of the velocity vectors, it can be written as \begin{align}
	\label{eq:OmegaComponents}
	u &= u' \cos(\theta) + \frac{\sin(\theta)}{\sqrt{1-w'^2}}[u' w' \cos(\phi) - v' \sin(\phi)] ,\nonumber \\
	v &= v' \cos(\theta) + \frac{\sin(\theta)}{\sqrt{1-w'^2}}[v' w' \cos(\phi) + u' \sin(\phi)] , \\
	w &= w' \cos(\theta) - \sqrt{1-w'^2} \sin(\theta) \cos(\phi) , \nonumber
\end{align} where \begin{align}
	\Omega_{-1} = \begin{bmatrix}
		u' & v' & w'
	\end{bmatrix}^{\mathrm{T}}, \; \; \; \Omega_{0} = \begin{bmatrix}
		u & v & w
	\end{bmatrix}^{\mathrm{T}} .
\end{align} We now want to compute the expectation value of $\Omega_0$, for which we need the probability density functions of the polar and azimuthal scattering angles $\theta$ and $\phi$. The cosine of the polar scattering angle $\cos(\theta) = \mu$ is a random variable for which the probability density function is given by the normalized differential scattering cross-section $\Sigma_s(E, \textbf{x}, \mu)$ \eqref{eq:DCS}. Usually, the polar scattering angle is dependent on the energy of the particle, but this dependence is neglected throughout the diffusive step. The azimuthal scattering angle $\phi$ is uniformly distributed between 0 and $2\pi$. We can now calculate the expectation value of equation \eqref{eq:OmegaComponents} is taken \begin{align}
	\expect{u|\Omega_{-1}} &= \expect{ u' \cos(\theta) + \frac{\sin(\theta)}{\sqrt{1-w'^2}}[u' w' \cos(\phi) - v' \sin(\phi)] |\Omega_{-1}} = u' \expect{\cos(\theta)}, \nonumber \\
	\expect{v|\Omega_{-1}} &= \expect{ v' \cos(\theta) + \frac{\sin(\theta)}{\sqrt{1-w'^2}}[v' w' \cos(\phi) + u' \sin(\phi)] |\Omega_{-1}} = v' \expect{\cos(\theta)}, \label{eq:ThisResult} \\
	\expect{w|\Omega_{-1}} &= \expect{ w' \cos(\theta) - \sqrt{1-w'^2} \sin(\theta) \cos(\theta) |\Omega_{-1}} = w' \expect{\cos(\theta)} , \nonumber
\end{align} since \begin{align}
	\label{eq:UniformAzimuthalScattering}
	\expect{\cos(\phi)} = \expect{\sin(\phi)} = 0 .
\end{align} In vector form, equation \eqref{eq:ThisResult} reads \begin{align}
	\label{eq:ThisResultVector}
	\expect{\Omega_{0}|\Omega_{-1}} = \Omega_{-1} \expect{\cos(\theta)}.     
\end{align}
In other words, the expected value of velocity $\Omega_{0}$ is along the direction of $\Omega_{-1}$. Considering the uniform azimuthal scattering angle $\phi$, this result is rather obvious. The result \eqref{eq:ThisResultVector} can now be used to obtain an expression for the expectation value of the next velocity in the chain $\Omega_1$. Using the total law of expectation, it follows that \begin{align}
	\expect{\Omega_1 | \Omega_{-1}} = \expect{\expect{\Omega_1|\Omega_0, \Omega_{-1}}| \Omega_{-1}} = \expect{\Omega_{0} \expect{\cos(\theta)} | \Omega_{-1}} = \Omega_0 \expect{\cos(\theta)}^2 ,
\end{align} such that in general \begin{align}
	\expect{\Omega_j | \Omega_{-1}} = \Omega_0 \expect{\cos(\theta)}^{j+1} .
\end{align} This result, combined with \eqref{eq:KDMC_dsk_exp}, can be used to simplify \eqref{eq:KDRmeantemp1}, which yields \begin{align}
	\expect*{\sum_{j=0}^{J} \Delta s_j \Omega_j | J, \Omega_{-1} } &= \frac{\Delta s}{J+1} \Omega_{-1} \expect{\cos(\theta)} \sum_{j=0}^{J} \expect{\cos(\theta)}^j \\
	&= \frac{\Delta s}{J+1} \Omega_{-1} \expect{\cos(\theta)} \frac{1 - \expect{\cos(\theta)}^{J+1} }{1 - \expect{\cos(\theta)}} \label{eq:KDRmeantemp5}.
\end{align} Now, continuing with equation \eqref{eq:KDRmeantemp2}, the mean of the kinetic motion becomes \begin{align}
	\label{eq:KDRmeantemp3}
	\expect{\Delta x | \Omega_{-1}} &= \frac{\Omega_{-1} \Delta s \expect{\cos(\theta)}}{1 - \expect{\cos(\theta)}} e^{-\Sigma_t \Delta s} \sum_{J=0}^{\infty} \frac{\left( \Sigma_t \Delta s \right)^J}{J!} \frac{1 - \expect{\cos(\theta)}^{J+1}}{J+1} . 
\end{align}
The infinite sum can be simplified using the power series of the exponential function. \begin{align}
	\sum_{J=0}^{\infty} \frac{\left( \Sigma_t \Delta s \right)^{J}}{J!} &\frac{1 - \expect{\cos(\theta)}^{J+1}}{J+1} \\ 
	&= \frac{1}{\Sigma_t \Delta s} \left[ \sum_{J=0}^{\infty} \frac{\left( \Sigma_t \Delta s \right)^{J+1}}{(J+1)!} - \sum_{J=0}^{\infty} \frac{\left( \Sigma_t \Delta s \expect{\cos(\theta)} \right)^{J+1}}{(J+1)!} \right]\\
	&= \frac{1}{\Sigma_t \Delta s} \left[ e^{\Sigma_t \Delta s}  - e^{\Sigma_t \Delta s \expect{\cos(\theta) }} \right].
\end{align} Finally, filling this result into equation \eqref{eq:KDRmeantemp3}, the formula for the mean of the kinetic motion taking into account the rotational dependencies \eqref{eq:KDRMean} is obtained.

\section{Pseudocode for particle simulation using KDR}
\label{section:kdrAlg}
Algorithm \ref{alg:KDR} implements a KDR simulation down a threshold energy $E_{th}$. The function SAMPLESTEPSIZE evaluates the scattering rate (cross-section) and samples an exponentially distributed random number. The function ENERGYLOSS computes the energy that a particle loses as it traverses the medium using equation \eqref{eq:computeEnergy}. The function ROTATEVELOCITY samples a polar and azimuthal scattering angle and returns a new velocity. The function SAMPLEMS samples the multiple scattering distribution and returns a new velocity. The function ROTATIONMATRIX constructs the rotation matrix \eqref{eq:RotateVelocity}, and INTERPOLATEVAR interpolates the lookup table for the variance. The boolean `useMS' is used to choose between the two solutions to sample a velocity after the KDR step.  
\begin{algorithm}
	\caption{KDR simulation down to threshold energy $E_{th}$. The algorithm is implemented in Python in \cite{willems_electrontransportcode_2023}.}
	\label{alg:KDR}
	\begin{algorithmic}[1]
		\Input{position $x_0$, velocity $\Omega_0$, energy $E_0$, step size $\Delta s$ and threshold energy $E_{th}$}
		\Output{position $x_k$, velocity $\Omega_k$ and energy $E_k$}
		\State $x_k \gets x_0$; $\Omega_k \gets \Omega_0$; $E_k \gets E_0$; $k \gets 0$
		\While{$E_k > E_{th}$}
		\State $\Delta s_k \gets$ SAMPLESTEPSIZE($x_k$, $E_k$)
		\State $\Delta E_k' \gets$ ENERGYLOSS($x_k$, $E_k$, $\Delta s_k$)
		\State $x_k' \gets x_k + \Delta s_k \Omega_k$ \Comment{Kinetic step}
		\State $E_k' \gets E_k - \Delta E_k' $
		\If{$E_k' > E_{th}$}  \Comment{Do diffusive step if there is energy left}
		\State $t_k \gets \Delta s - (\Delta s_k \;  \; \text{MOD} \; \; \Delta s) $
		\State $A_k \gets \text{eq. \eqref{eq:KDMCAdvectionCoef}}$
		\State $V \gets $ INTERPOLATEVAR($E_k'$, $t_k$, $\rho$)
		\State $R \gets $ ROTATIONMATRIX($\Omega_k$)
		\State $\xi_k \gets \mathcal{N}(\mathbf{0}, I_3)$
		\State $x_{k+1} \gets x_k' + A_k t_k + R \sqrt{V} \xi_k$ 
		\State $E_{k+1} \gets  E_k' -$ ENERGYLOSS($x_k'$, $E_k'$, $t_k$)
		\If{useMS} \Comment{Sample multiple scattering distribution}
		\State $\Omega_{k+1} \gets$ SAMPLEMS($E_k'$, $t_k$, $\rho$, $\Omega_k$)
		\Else 
		\State $\Omega' \gets \frac{\mathbf{x}_{k+1} - \mathbf{x}_k'}{| \mathbf{x}_{k+1} - \mathbf{x}_k'|}$
		\State $\Omega_{k+1} \gets $ ROTATEVELOCITY($x_{k+1}$, $E_{k+1}$, $\Omega'$)
		\EndIf
		\EndIf
		\State $k \gets k + 1$
		\EndWhile
	\end{algorithmic}
\end{algorithm}

\end{appendices}

\newpage
\bibliographystyle{acm}
\bibliography{bibliography, masterThesis}

\end{document}